# Work-Energy Principle Based Characteristic Mode Theory for Yagi-Uda Array Antennas

Ren-Zun Lian, Ming-Yao Xia, *Senior Member, IEEE*, and Xing-Yue Guo, *Student Member, IEEE*

***Abstract*—Work-energy principle (WEP) governing the work-energy transformation process of Yagi-Uda array antennas is derived. Driving power as the source to sustain a steady work-energy transformation is introduced. Employing WEP and driving power, the essential difference between the working mechanisms of scattering objects and Yagi-Uda *array antennas* is revealed. The difference exposes that the conventional characteristic mode theory (CMT) for scattering objects cannot be directly applied to Yagi-Uda *array antennas*. Under WEP framework, this paper proposes a generalized CMT for Yagi-Uda antennas. By orthogonalizing driving power operator (DPO), the WEP-based CMT can construct a set of energy-decoupled characteristic modes (CMs) for an objective Yagi-Uda antenna, and then can provide an effective modal analysis for the Yagi-Uda antenna. In addition, a uniform interpretation for the physical meaning of the characteristic values / modal significances (MSs) of metallic, material, and metal-material composite Yagi-Uda antennas is also obtained by employing the WEP-based modal decomposition and the field-current interaction expression of driving power.**

## I. Introduction

YAGI-UDA array antenna was first studied by Uda and Yagi in the early 1920s, and publicly reported in the middle 1920s [1]-[2]. In 1984, the *Proceedings of the IEEE* reprinted several classical articles for celebrating the centennial year of IEEE (1884-1984), and Yagi's article [2] became the only reprinted one in the realm of electromagnetic (EM) antenna. This fact clearly illustrates the great significance of Yagi-Uda antenna in *Antennas & Propagation Society*. Some histories about Yagi-Uda antenna can be found in [3]-[5].

A classical metallic Yagi-Uda antenna [6] is shown in Fig. 1, and it is constituted by a row of discrete metallic linear elements, one of which is driven by a voltage source while the others act as parasitic radiators whose currents are induced by near-field mutual coupling [7]-[10]. The linear metallic Yagi-Uda antenna is a typical discrete-element travelling-wave end-fire antenna, which usually works at HF (3-30 MHz), VHF (30-300 MHz) and UHF (300-3000 MHz) etc. bands [7]-[10].

R.-Z. Lian, M.-Y. Xia and X.-Y. Guo are with the Department of Electronics, School of Electronics Engineering and Computer Science, Peking University, Beijing 100871, China. (e-mail: rzlian@vip.163.com; myxia@pku.edu.cn).

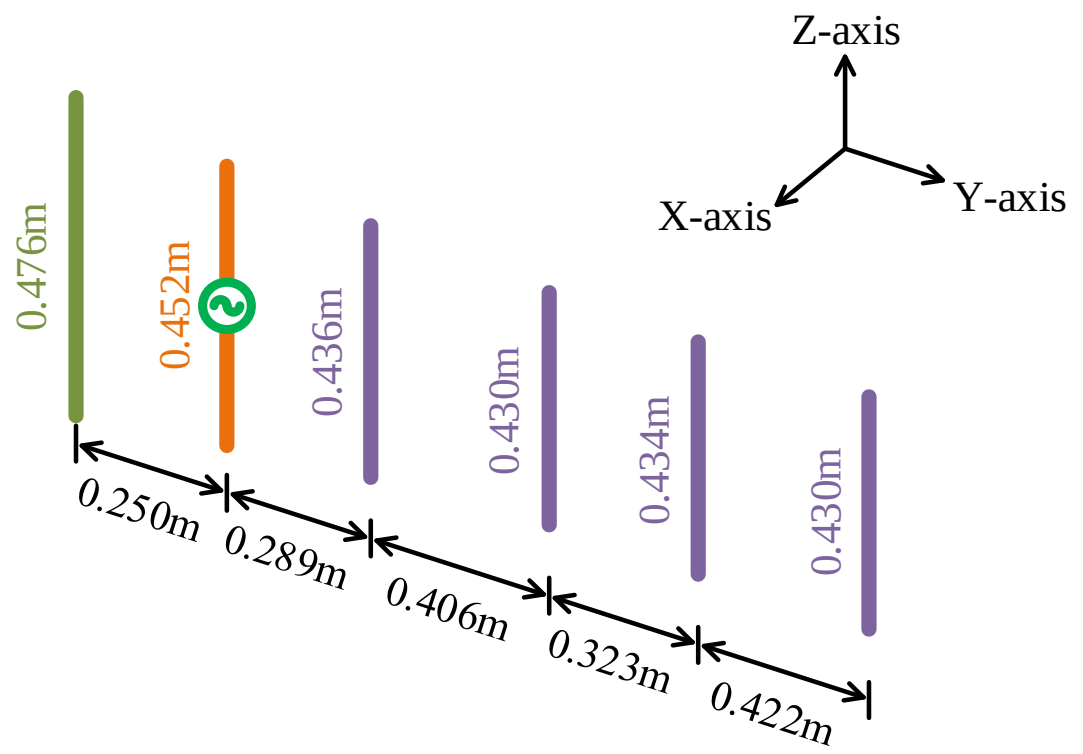

Fig. 1. Geometry and size of a typical 6-element linear metallic Yagi-Uda array antenna reported in [6]. The dominant resonant mode of the Yagi-Uda antenna works at 300 MHz (calculated from the formulation proposed in [6]).

Besides the most classical linear metallic Yagi-Uda antenna shown in Fig. 1, there also exist many different variants (sometimes called quasi Yagi-Uda antennas). The quasi Yagi-Uda antennas have been widely applied in the applications of frequency-modulated broadcast [11], domestic/mobile television signal transmission [12]-[14], point-to-point communication [15]-[19], long-distance communication [20]-[24], mobile communication [25]-[29], wireless local area network [30]-[34] and radio frequency identification [35]-[38] etc. due to their typical features of high radiation efficiency, narrow beamwidth / high gain and directivity, high front-to-back ratio, low level of minor lobes, good cross-polar discrimination, reasonable bandwidth, low cost and ease of fabrication etc. [7]-[10].

According to the difference of their constituent components, the various Yagi-Uda antennas can be categorized into three classes — metallic Yagi-Uda antennas [6], [39]-[41], material Yagi-Uda antennas [42]-[43], and metal-material composite Yagi-Uda antennas [12]-[38]. The one shown in Fig. 1 is just a typical 6-element metallic Yagi-Uda antenna, which had been carefully studied in [6] based on analytical method, and it has a dominant resonant mode working at 300 MHz (calculated from the formulation proposed in [6]), and the resonant mode has the far-field radiation pattern shown in Fig. 2, which is end-fire.

The analysis and design for resonant modes are the important topics in the realm of antenna engineering, and there have been some classical modal analysis and design theories, such as model-based modal theories (e.g., cavity model theory [44]-[45] and dielectric waveguide model theory [46]-[47]), eigen-mode theory [48]-[49], and characteristic mode theory (CMT) [50]-[55] etc. Among the various theories, the CMT has been attached great importance in the realm of antenna engineering [54] recently, because the CMT not only has a very wide applicable range but also is very easy for numerical realization. But unfortunately, the conventional CMT cannot be directly applied to doing the modal analysis for Yagi-Uda *array antenna*s as exhibited below.

By directly applying the conventional CMT to the Yagi-Uda *antenna* in Fig. 1, the modal significances (MSs) associated to the obtained characteristic modes (CMs) are shown in Fig. 3, and the far-field power-density distributions of the resonant CMs are illustrated in Fig. 4. Evidently, both the resonance frequencies and far-field power-density distributions are not consistent with the one shown in Fig. 2. The reasons leading to this inconsistency mainly originate from the following ones.

- The working mechanisms of scattering objects and transmitting antennas are different from each other. Specifically,
  - as shown in Fig. 5(a), scattering object is a structure which is under the illumination of an externally incident field and generates a secondary scattered field;
  - as shown in Fig. 5(b), transmitting antenna is *a device used for transmitting electromagnetic signals or power* [56, pp. 41].
- Yagi-Uda arrays belong to the family of transmitting antennas rather than the family of scattering objects.
- The conventional CMT is a modal analysis method for scattering objects rather than for transmitting antennas (though it indeed works for some very special antennas, such as voltage-source-driven *single* dipole), and this conclusion will be carefully explained in Sec. II-B.

More detailed discussions for the reasons will be given in Sec. II from the aspects of energy source (Sec. II-A), work-energy transformation process (Sec. II-A), CM calculation process (Sec. II-B), and resonant current distribution (Sec. II-D).

Above these clearly expose the significance and value to generalize the conventional CMT for scattering objects to a generalized CMT for Yagi-Uda antennas, and this paper is devoted to doing the generalization by employing the work-energy principle (WEP) governing the work-energy transformation process of Yagi-Uda antennas. This paper is organized as follows: Sec. II discusses the WEP-based CMT (WEP-CMT) for metallic Yagi-Uda antennas; Sec. III generalizes the WEP-CMT to material Yagi-Uda antennas; Sec. IV further generalizes the WEP-CMT to composite Yagi-Uda antennas; Sec. V concludes this paper; some detailed formulations related to this paper are provided in the appendices.

In what follows, the $e^{j\omega t}$ convention and inner product $<\boldsymbol{f},\boldsymbol{g}>_{\Omega}=\int_{\Omega}\boldsymbol{f}^{\dagger}\cdot\boldsymbol{g}d\Omega$ are used throughout, where superscript "$\dagger$" is the conjugate transpose operation for a scalar/vector/matrix. The environment surrounding Yagi-Uda antenna is the free space with $\mu_0$ and $\varepsilon_0$. Time-domain and frequency-domain "field, surface/volume current" are denoted as "$\boldsymbol{\mathcal{F}},\boldsymbol{\mathcal{J}}/\boldsymbol{\acute{j}}$" and "$\boldsymbol{F},\boldsymbol{J}/\boldsymbol{j}$" respectively; frequency-domain power quadratic matrix and surface/volume current expansion coefficient vector are denoted as $\mathbb{P}$ and $\mathbb{J}/\mathbb{j}$ respectively. In addition, for the linear quantities (e.g., electric field intensity), we have that $\boldsymbol{\mathcal{E}}=\text{Re}\{\boldsymbol{E}e^{j\omega t}\}$; for the power-type quadratic quantity, we have that $\text{Re}\{(1/2)\boldsymbol{J}^{\dagger}\cdot\boldsymbol{E}\}=(1/T)\int_0^T\boldsymbol{\mathcal{J}}\cdot\boldsymbol{\mathcal{E}}dt$, where $T$ is the time period of the time-harmonic EM field.

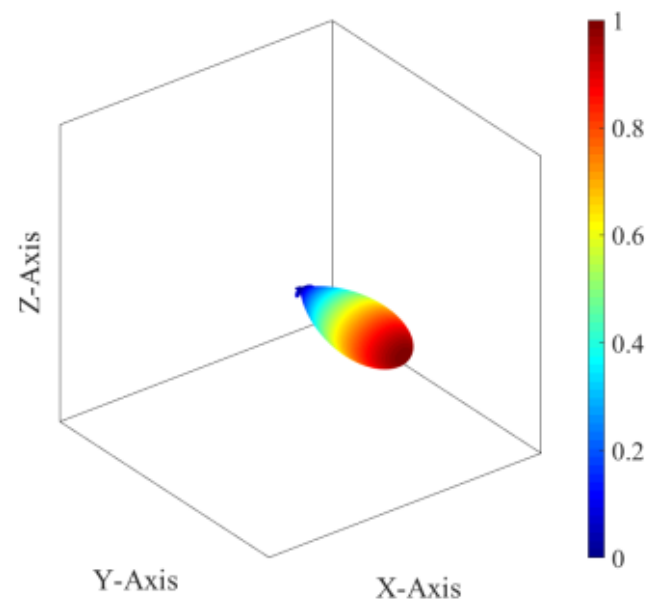

Fig. 2. End-fire radiation pattern of the dominant resonant mode (working at 300 MHz) of the metallic Yagi-Uda array antenna shown in Fig. 1.

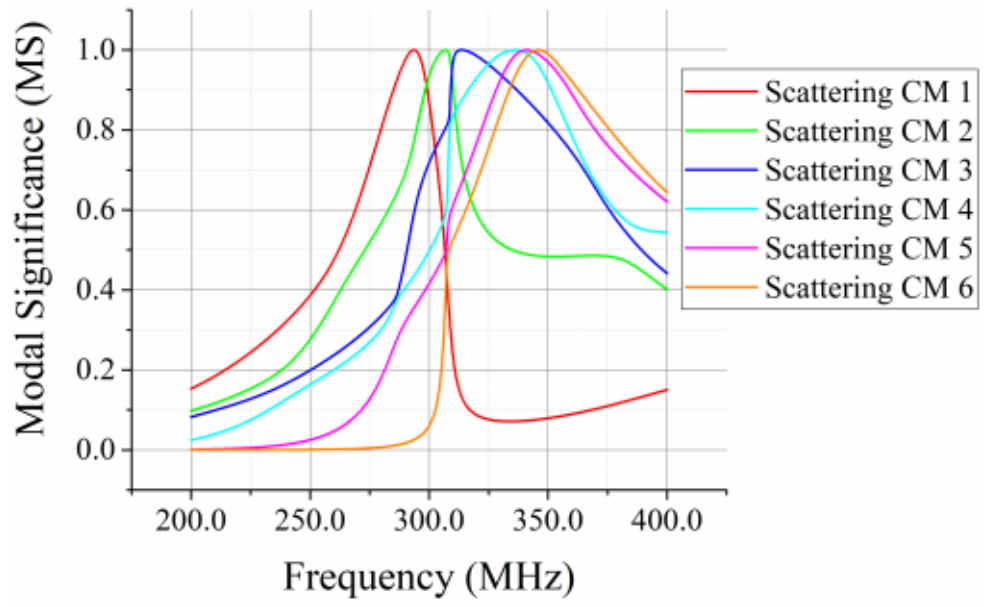

Fig. 3. MSs associated to the first six lower-order scattering CMs calculated from the conventional CMT established by Harrington *et al.* in [52].

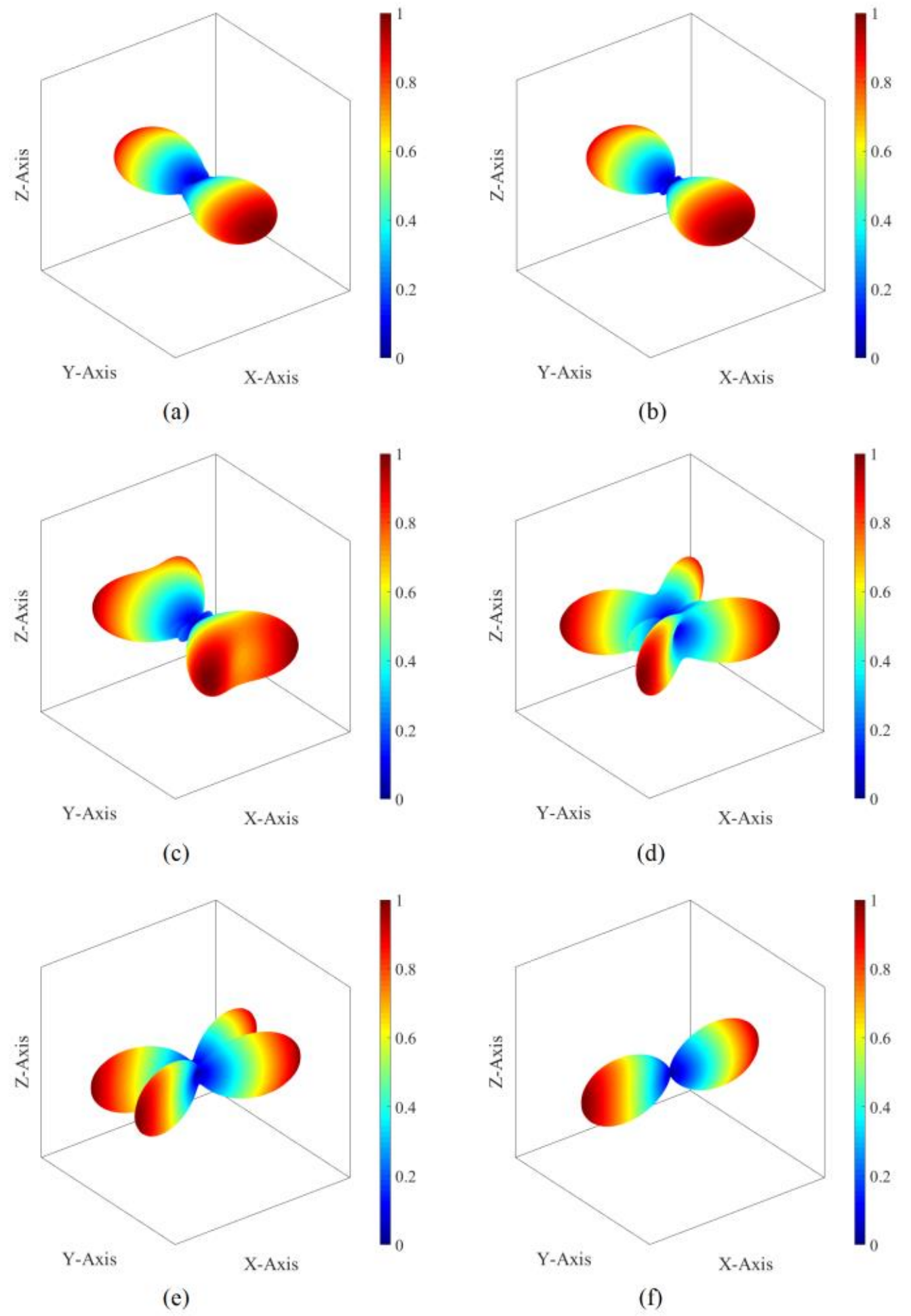

Fig. 4. Far-field power-density distributions of the resonant scattering (a) CM 1 at 293.6 MHz, (b) CM 2 at 307.0 MHz, (c) CM3 at 313.6 MHz, (d) CM4 at 336.2 MHz, (e) CM5 at 340.9 MHz, and (f) CM6 at 345.9 MHz. Here, the scattering CMs 1 ~ 6 correspond to the ones shown in Fig. 3.

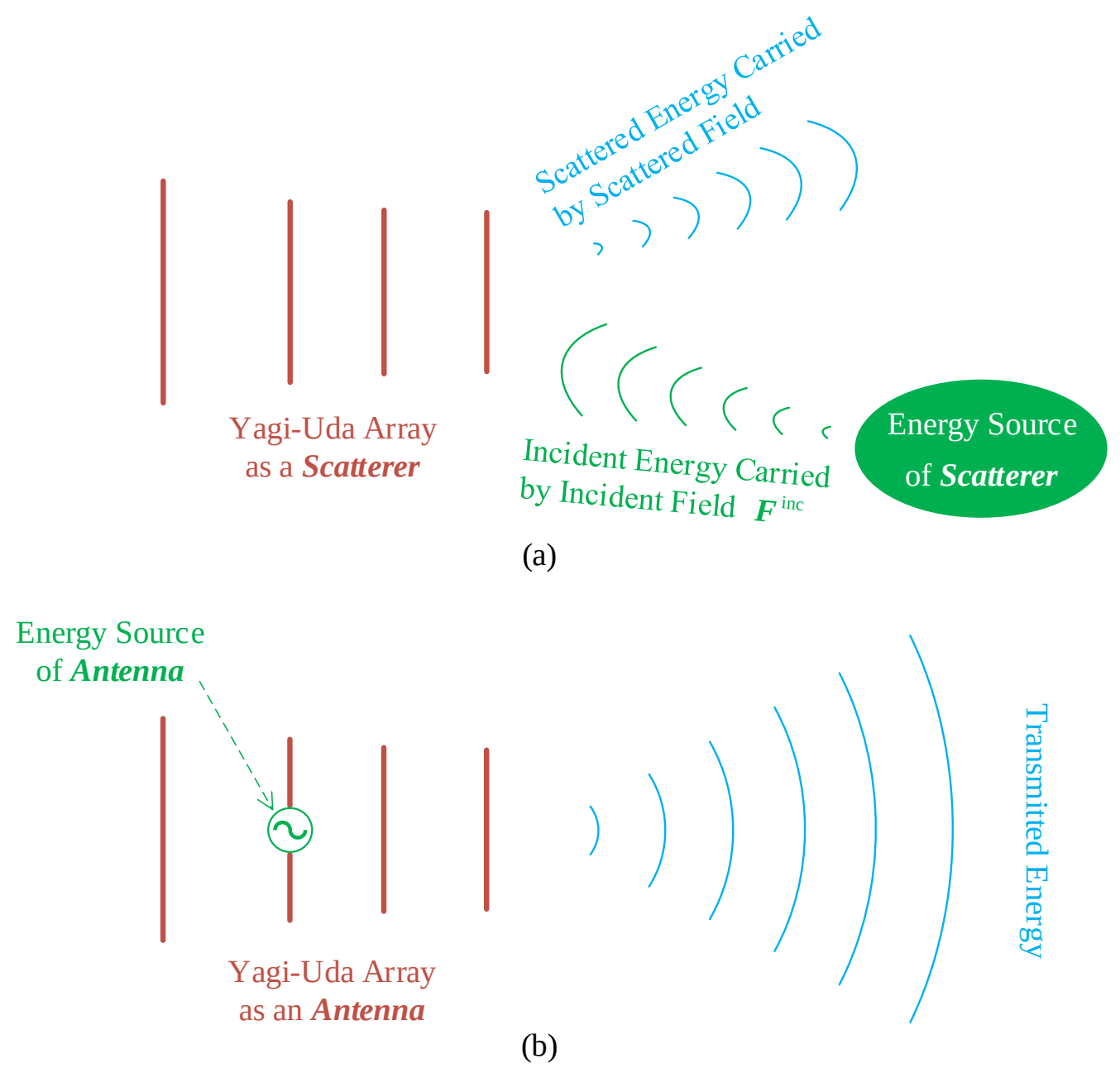

Fig. 5. (a) External-incident-field-driven Yagi-Uda array ***scatterer*** and (b) internal-voltage-source-driven Yagi-Uda array ***antenna***.

## II. WEP-CMT for Metallic Yagi-Uda Array Antennas

Taking the one shown in Fig. 1 as a typical example, this section proposes a generalized WEP-CMT for metallic Yagi-Uda *array antennas*.

### A. Working Mechanism of Metallic Yagi-Uda Array Antennas — Focusing on Energy Source and Work-Energy Transformation Process

Traditionally, all the elements in a Yagi-Uda antenna are divided into three groups — feeding element, reflecting element, and directing elements [7]-[10]. In fact, all the elements can also be alternatively divided into two groups — active element and passive/parasitic elements [20], where the former is just the feeding element and the latter is the union of reflecting element and directing elements as shown in Fig. 6. For highlighting the working mechanism of the Yagi-Uda antenna and simplifying the mathematical formulation of following discussion, the second grouping way is selected in this Sec. II.

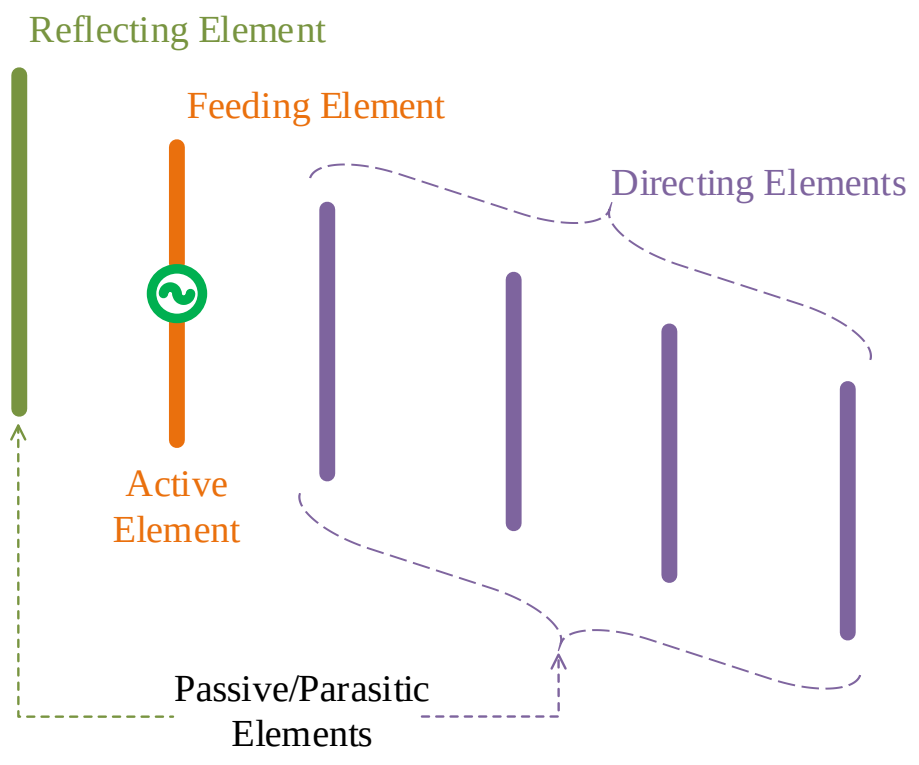

Fig. 6. Two different grouping ways for the elements used to constitute the Yagi-Uda array antenna shown in Fig. 1.

As shown in Fig. 6, the steady energy output of the Yagi-Uda antenna is sustained by a voltage source which drives the active element. The voltage driving has a field effect, i.e., the voltage driving can be equivalently viewed as a field driving. The voltage driving acts on the active element directly (but doesn't act on the passive elements directly), so the equivalent field driving acts on the active element only (but doesn't act on the passive elements), i.e., the driving field is localized/restricted in the region occupied by the active element.

The action of the driving field on active element will induce a current on the active element, and the current will generate a field on surrounding environment. Similarly, the field generated by active element will act on passive elements, and the action will lead to some induced currents on the passive elements, and the currents will generate some fields on surrounding environment. In fact, there also exists a reaction from the fields generated by passive elements to the current distributing on active element, and the reaction will affect the active current distribution. Through a complicated process, the above actions and reactions will reach a dynamic equilibrium finally, because the EM problem considered here is time-harmonic.

For convenience of the following discussions, the boundary surfaces of active element and passive elements are denoted as $S_{\rm a}$ and $S_{\rm p}$ respectively; the three-dimensional Euclidean space is denoted as $\mathbb{E}^3$; the boundary of $\mathbb{E}^3$ is denoted as $S_\infty$, which is a closed spherical surface with infinite radius. The driving field generated by voltage driver is denoted as $\boldsymbol{F}_{\rm driv}$. At the state of dynamic equilibrium, the currents distributing on $S_{\rm a}$ and $S_{\rm p}$ are denoted as $\boldsymbol{J}_{\rm a}$ and $\boldsymbol{J}_{\rm p}$ respectively. The fields generated by $\boldsymbol{J}_{\rm a}$ and $\boldsymbol{J}_{\rm p}$ are denoted as $\boldsymbol{F}_{\rm a}$ and $\boldsymbol{F}_{\rm p}$ respectively, and the summation of $\boldsymbol{F}_{\rm a}$ and $\boldsymbol{F}_{\rm p}$ is denoted as $\boldsymbol{F}$, i.e., $\boldsymbol{F} = \boldsymbol{F}_{\rm a} + \boldsymbol{F}_{\rm p}$, and $\boldsymbol{F}$ is just the field generated by all elements.

The *conservation law of energy* tells us that the above actions and reactions among voltage driver, active element, and passive elements will result in a work-energy transformation, and the work-energy transformation can be quantitatively expressed as the following work-energy principle (WEP)

$$\overbrace{(1/2)\langle \boldsymbol{J}_{\rm a}, \boldsymbol{E}_{\rm driv}\rangle_{S_{\rm a}}}^{P_{\rm driv}} = (1/2)\oint\!\!\!\oint_{S_\infty} (\boldsymbol{E}\times\boldsymbol{H}^\dagger)\cdot d\boldsymbol{S} + j2\omega\Big[(1/4)\langle \boldsymbol{H},\boldsymbol{B}\rangle_{\mathbb{E}^3} - (1/4)\langle \boldsymbol{D},\boldsymbol{E}\rangle_{\mathbb{E}^3}\Big] \tag{1}$$

where $P_{\rm driv} = (1/2) < \boldsymbol{J}_{\rm a}, \boldsymbol{E}_{\rm driv} >_{S_{\rm a}}$ and $\boldsymbol{D} = \varepsilon_0 \boldsymbol{E}$ and $\boldsymbol{B} = \mu_0 \boldsymbol{H}$. A rigorous mathematical derivation for (1) is given in App. A.

Above WEP clearly exhibits the work-energy transformation process of the Yagi-Uda array antenna: the voltage source drives whole Yagi-Uda array by directly acting on the active element, and the driving power $P_{\rm driv}$ (i.e., the power done by $\boldsymbol{E}_{\rm driv}$ on $\boldsymbol{J}_{\rm a}$) is finally transformed into two parts, where a part is radiated to $S_\infty$ and another part is reactively stored in $\mathbb{E}^3$.

Obviously, the above-mentioned working mechanism of metallic Yagi-Uda *array antennas* is very different from the working mechanism of scattering objects (such as the metallic Yagi-Uda *array scatterer* shown in Fig. 5(a)), and the difference is mainly reflected on their different energy sources and different work-energy transformation processes.

- The energy source of scattering object is an externally incident field as shown in Fig. 5(a). When the whole Yagi-Uda array is treated as a scattering object, the driving power can be expressed as field-current interaction $(1/2) < \boldsymbol{J}_{\text{a}} + \boldsymbol{J}_{\text{p}}, \boldsymbol{E}^{\text{inc}} >_{S_{\text{a}} \cup S_{\text{p}}}$ where $\boldsymbol{E}^{\text{inc}}$ is just the externally incident field as shown in Fig. 5(a) [55], and the driving field (i.e., the incident field $\boldsymbol{E}^{\text{inc}}$) supplies energies to active element and passive elements *simultaneously*;
- The energy source of Yagi-Uda array antenna is a localized driving field (i.e., the equivalent field effect of the voltage source shown in Figs. 5(b) and 6), and the driving field is restricted in a finite region (e.g., the active element shown in Fig. 6). When the Yagi-Uda array is treated as a transmitting antenna, the driving power $P_{\text{driv}}$ is expressed as the one given in (1), and the driving field $\boldsymbol{E}_{\text{driv}}$ directly supplies energy to active element only, and the energy supplied to passive elements originates from the near-field mutual coupling between the active element and passive elements [7]-[10], i.e., the active element acts as an *energy relay* between the voltage source and the passive elements.

In the language of the *IEEE standards terms* [56], the above difference between the working mechanisms of scattering object and Yagi-Uda array antenna is stated as follows:

- scattering object, such as the Yagi-Uda array scatterer shown in Fig. 5(a), is *a secondary structure generating scattered fields resulted from the scattered currents induced on the structure by some fields incident on the structure from some primary sources* [56, pp. 1006];
- Yagi-Uda array antenna, such as the one shown in Fig. 5(b), is *a device that generates high-frequency electric energy, controlled or modulated, which can be emitted from a finite region in the form of unguided waves* [34, pp. 369 & 1210].

Based on the above-exposed difference between the working mechanisms of scattering object and Yagi-Uda array antenna, we generalize the conventional CMT for the former to a generalized CMT for the latter in the following Sec. II-B.

### *B. Core Physical Features and Mathematical Calculation Processes of Characteristic Modes*

The WEP-CMT proposed here focuses on constructing a set of CMs satisfying the following power-decoupling relation

$$\left(1/2\right)\left\langle \boldsymbol{J}_{\text{a};\xi}, \boldsymbol{E}_{\text{driv};\zeta} \right\rangle_{S_{\text{a}}} = \delta_{\xi\zeta}\left(1 + j\lambda_{\xi}\right) \tag{2}$$

and then satisfying the following energy-decoupling relation, i.e., time-average power-decoupling relation,

$$\left(1/T\right)\int_{t_0}^{t_0+T} \left\langle \mathcal{J}_{\text{a};\xi}, \mathcal{E}_{\text{driv};\zeta} \right\rangle_{S_{\text{a}}} dt = \delta_{\xi\zeta} \tag{3}$$

where $\delta_{\xi\zeta}$ is Kronecker's delta symbol, and $\lambda_{\xi}$ is characteristic value, and the real parts of modal complex powers have been normalized to 1 just like the conventional scatterer-oriented CMT [52]-[54] did. Thus, there exists Parseval's identity

$$\left(1/T\right)\int_{t_0}^{t_0+T} \left\langle \mathcal{J}_{\text{a}}, \mathcal{E}_{\text{driv}} \right\rangle_{S_{\text{a}}} dt = \sum_{\xi}\left|c_{\xi}\right|^2 \tag{4}$$

in which $c_{\xi}$ is the CM-based modal expansion coefficient, and $c_{\xi} = (1/2) < \boldsymbol{J}_{\text{a};\xi}, \boldsymbol{E}_{\text{driv}} >_{S_{\text{a}}} \big/ (1 + j\lambda_{\xi})$ where $\boldsymbol{E}_{\text{driv}}$ is the frequency-domain version of a previously known time-domain driving field $\mathcal{E}_{\text{driv}}$, which uniquely determines the current $\mathcal{J}_{\text{a}}$ due to *EM field unique theorem* [10], [57]. The above (3) and (4) have a very clear physical interpretation: the CMs constructed above don't have net energy exchange in any integral period.

In fact, besides the power-decoupling relation (2) and energy-decoupling relation (3), the CMs also satisfy the following radiated power orthogonality and reactive power orthogonality

$$\begin{aligned}\delta_{\xi\zeta} &= \left(1/2\right)\oiint_{S_\infty}\left(\boldsymbol{E}_{\zeta} \times \boldsymbol{H}_{\xi}^{\dagger}\right)\cdot d\boldsymbol{S} \\ &= \left(1/2\eta_0\right)\left\langle \boldsymbol{E}_{\xi}, \boldsymbol{E}_{\zeta}\right\rangle_{S_\infty} = \left(\eta_0/2\right)\left\langle \boldsymbol{H}_{\xi}, \boldsymbol{H}_{\zeta}\right\rangle_{S_\infty}\end{aligned} \tag{5}$$

$$\lambda_{\xi}\ \delta_{\xi\zeta} = 2\omega\left[\left(1/4\right)\left\langle \boldsymbol{H}_{\xi}, \boldsymbol{B}_{\zeta}\right\rangle_{\mathbb{E}^3} - \left(1/4\right)\left\langle \boldsymbol{D}_{\xi}, \boldsymbol{E}_{\zeta}\right\rangle_{\mathbb{E}^3}\right] \tag{6}$$

as rigorously proved in App. C, where the second and third equalities in (5) are originated from the Sommerfeld's radiation condition $\lim_{r \to \text{infinity}}(\nabla \times \boldsymbol{F} + jk_0\hat{\boldsymbol{n}}_\infty \times \boldsymbol{F}) = 0$ at infinity (where $k_0 = \omega\sqrt{\mu_0\varepsilon_0}$, and $\hat{\boldsymbol{n}}_\infty$ is the outer normal direction of $S_\infty$) [57, Sec. 3-5] and the homogeneous Maxwell's equations $\nabla \times \boldsymbol{H} = j\omega\varepsilon_0\boldsymbol{E}$ and $\nabla \times \boldsymbol{E} = -j\omega\mu_0\boldsymbol{H}$ at infinity, and $\eta_0 = \sqrt{\mu_0/\varepsilon_0}$ is free-space wave impedance.

Besides the previously exhibited field-current interaction form $(1/2) < \boldsymbol{J}_{\text{a}}, \boldsymbol{E}_{\text{driv}} >_{S_{\text{a}}}$, the driving power $P_{\text{driv}}$ also has the following integral operator expression

$$P_{\text{driv}} = \left(1/2\right)\left\langle \boldsymbol{J}_{\text{a}}\ ,\ j\omega\mu_0\mathcal{L}_0\left(\boldsymbol{J}_{\text{a}} + \boldsymbol{J}_{\text{p}}\right)\right\rangle_{S_{\text{a}}} \tag{7}$$

called driving power operator (DPO), in which the operator $\mathcal{L}_0$ is as $\mathcal{L}_0(\boldsymbol{X}) = [1 + (1/k_0^2)\nabla\nabla\cdot]\int_\Omega G_0(\boldsymbol{r}, \boldsymbol{r}')\boldsymbol{X}(\boldsymbol{r}')d\Omega'$ and the scalar Green's function is as $G_0(\boldsymbol{r}, \boldsymbol{r}') = e^{-jk_0|\boldsymbol{r}-\boldsymbol{r}'|}/4\pi|\boldsymbol{r}-\boldsymbol{r}'|$. In addition, the above integral form (7) of DPO can be easily discretized into the following matrix form

$$P_{\text{driv}} = \mathbb{J}_{\text{a}}^{\dagger}\cdot\begin{bmatrix}\mathbb{P}_{\text{aa}} & \mathbb{P}_{\text{ap}}\end{bmatrix}\cdot\begin{bmatrix}\mathbb{J}_{\text{a}} \\ \mathbb{J}_{\text{p}}\end{bmatrix} \tag{8}$$

where $\mathbb{J}_{\text{a}}$ and $\mathbb{J}_{\text{p}}$ are the basis function expansion coefficient vectors of $\boldsymbol{J}_{\text{a}}$ and $\boldsymbol{J}_{\text{p}}$ respectively, and the formulations for calculating the sub-matrices $\mathbb{P}_{\text{aa}}$ and $\mathbb{P}_{\text{ap}}$ are given in App. B.

In fact, the $\mathbb{J}_{\text{a}}$ and $\mathbb{J}_{\text{p}}$ are not independent, and they satisfy the following transformation relation

$$\mathbb{J}_{\text{p}} = \mathbb{T}\cdot\mathbb{J}_{\text{a}} \tag{9}$$

which originates from the tangential electric field boundary condition on $S_{\text{p}}$. The formulas for calculating $\mathbb{T}$ are given in App. B. Substituting (9) into (8), it is easy to derive following

$$P_{\text{driv}} = \mathbb{J}_{\text{a}}^{\dagger}\cdot\mathbb{P}_{\text{driv}}\cdot\mathbb{J}_{\text{a}} \tag{10}$$

with independent current $\mathbb{J}_{\text{a}}$ only, where the formulations for calculating $\mathbb{P}_{\text{driv}}$ are also given in App. B.

The energy-decoupled CMs satisfying decoupling relations (2)&(3) and orthogonality relations (5)&(6) can be derived from solving the following characteristic equation

$$\mathrm{P}_{\mathrm{driv}}^{-}\cdot \mathrm{J}_{\mathrm{a}} = \lambda\, \mathrm{P}_{\mathrm{driv}}^{+}\cdot \mathrm{J}_{\mathrm{a}} \tag{11}$$

where $\mathrm{P}_{\mathrm{driv}}^{+} = [\mathrm{P}_{\mathrm{driv}} + \mathrm{P}_{\mathrm{driv}}^{\dagger}]/2$ and $\mathrm{P}_{\mathrm{driv}}^{-} = [\mathrm{P}_{\mathrm{driv}} - \mathrm{P}_{\mathrm{driv}}^{\dagger}]/2j$.

In the previous Sec. II-A, the difference between scattering object and Yagi-Uda array antenna are carefully explained in the aspects of energy source and work-energy transformation process. Here, the difference is further exposed from the aspect of CM calculation process as follows:

- when the whole Yagi-Uda array is treated as a scattering object, currents $\mathrm{J}_{\mathrm{a}}$ and $\mathrm{J}_{\mathrm{p}}$ are excited independently, so they are independent of each other, and thus there doesn't exist a definite transformation relation between $\mathrm{J}_{\mathrm{a}}$ and $\mathrm{J}_{\mathrm{p}}$;
- when the Yagi-Uda array is treated as a transmitting antenna, the currents $\mathrm{J}_{\mathrm{a}}$ and $\mathrm{J}_{\mathrm{p}}$ are not independent as explained in Sec. II-A (physically) and App. B (mathematically), so the dependent current must be eliminated from (8). Otherwise, some spurious modes will be outputted from the characteristic equation, and some similar explanations for the spurious mode problem can be found in [55] and [58].

In fact, the above this is just the reason why the scattering CMs shown in Fig. 4 are inconsistent with the transmitting CM shown in Fig. 2. In the future Sec. II-D, we will further discuss the inconsistence from the aspect of modal current distribution.

It is necessary to emphasize that: for a lumped-port-driven antenna which is constituted by a *single* metal, such as a voltage-source-driven *single* dipole, its scattering CMs are the same as its transmitting CMs, because its scattering and transmitting CM formulas are the same in this special case. But, this conclusion is not always correct, especially for the lumped-port-driven *array* antennas (such as Yagi-Uda antenna) and the wave-port-fed antennas (such as horn antenna).

### *C. Physical Meanings of Characteristic Value and Modal Significance*

In the case of scattering objects, the physical meaning of characteristic value (CV) had been studied by many scholars. [52] provided a physical interpretation for the CV of metallic scatterer based on Poynting's theorem. [53, Sec. II] tried to generalize the Poynting's theorem based interpretation to the CV of material scatterer, but didn't succeed. It was found out in [59]-[60] that the Poynting's theorem based interpretation for metallic CV is not suitable for material CV, and then also not suitable for the CV of metal-material composite scatterer.

In fact, besides the CV of various scatterers, the physical meaning of the CVs of metallic, material, and composite Yagi-Uda antennas can also not be uniformly interpreted by Poynting's theorem. Taking metallic Yagi-Uda antenna as an example, we propose an alternative physical interpretation for CV by using the concept of driving power and the viewpoint of modal decomposition [61] introduced under WEP framework, and the interpretation can be easily generalized from metallic Yagi-Uda antenna to material and composite Yagi-Uda antennas, because the driving power and modal decomposition are universally applicable to the various Yagi-Uda antennas.

The power-based modal decomposition given in [61] can be directly generalized to the metallic Yagi-Uda antenna, and it decomposes any working mode $\mathrm{J}_{\mathrm{a}}$ in terms of the summation of three energy-decoupled fundamental components as that $\mathrm{J}_{\mathrm{a}} = \mathrm{J}_{\mathrm{a}}^{\mathrm{ind}} + \mathrm{J}_{\mathrm{a}}^{\mathrm{res}} + \mathrm{J}_{\mathrm{a}}^{\mathrm{cap}}$. Here, $\mathrm{J}_{\mathrm{a}}^{\mathrm{ind/res/cap}} = \sum_{i/r/c} c_{i/r/c} \mathrm{J}_{\mathrm{a};i/r/c}^{\mathrm{ind/res/cap}}$ is constituted by all inductive/resonant/capacitive CMs. Then, the current $\boldsymbol{J}_{\mathrm{a}}$ and field $\boldsymbol{E}_{\mathrm{driv}}$ are also correspondingly decomposed into three fundamental energy-decoupled components as that $\boldsymbol{J}_{\mathrm{a}} = \boldsymbol{J}_{\mathrm{a}}^{\mathrm{ind}} + \boldsymbol{J}_{\mathrm{a}}^{\mathrm{res}} + \boldsymbol{J}_{\mathrm{a}}^{\mathrm{cap}}$ and $\boldsymbol{E}_{\mathrm{driv}} = \boldsymbol{E}_{\mathrm{driv}}^{\mathrm{ind}} + \boldsymbol{E}_{\mathrm{driv}}^{\mathrm{res}} + \boldsymbol{E}_{\mathrm{driv}}^{\mathrm{cap}}$.

Based on the above modal decomposition, we can define the following Θ-factor

$$\Theta(\mathrm{J}_{\mathrm{a}}) = \frac{\overbrace{(\mathrm{J}_{\mathrm{a}}^{\mathrm{ind}})^{\dagger}\cdot \mathrm{P}_{\mathrm{driv}}^{-}\cdot \mathrm{J}_{\mathrm{a}}^{\mathrm{ind}}}^{\text{this term is positive}} - \overbrace{(\mathrm{J}_{\mathrm{a}}^{\mathrm{cap}})^{\dagger}\cdot \mathrm{P}_{\mathrm{driv}}^{-}\cdot \mathrm{J}_{\mathrm{a}}^{\mathrm{cap}}}^{\text{this term is negative}}}{\underbrace{\mathrm{J}_{\mathrm{a}}^{\dagger}\cdot \mathrm{P}_{\mathrm{driv}}^{+}\cdot \mathrm{J}_{\mathrm{a}}}_{\text{this term is positive}}} \tag{12}$$

for any working mode $\mathrm{J}_{\mathrm{a}}$. In the numerator of the right-hand side of (12), the positive term originates from that the phase of field component $\boldsymbol{E}_{\mathrm{driv}}^{\mathrm{ind}}$ is ahead of the phase of current component $\boldsymbol{J}_{\mathrm{a}}^{\mathrm{ind}}$, and the negative term originates from that the phase of field component $\boldsymbol{E}_{\mathrm{driv}}^{\mathrm{cap}}$ lags behind the phase of current component $\boldsymbol{J}_{\mathrm{a}}^{\mathrm{cap}}$. In addition, field component $\boldsymbol{E}_{\mathrm{driv}}^{\mathrm{ind}}/\boldsymbol{E}_{\mathrm{driv}}^{\mathrm{cap}}$ and current component $\boldsymbol{J}_{\mathrm{a}}^{\mathrm{cap}}/\boldsymbol{J}_{\mathrm{a}}^{\mathrm{ind}}$ are always energy-decoupled due to the energy-decoupling feature (2) of CMs. Thus, the larger $\Theta(\mathrm{J}_{\mathrm{a}})$ is, the stronger field-current phase-mismatching is. Then, the $\Theta(\mathrm{J}_{\mathrm{a}})$ quantitatively characterizes the mismatching degree between the phase of field $\boldsymbol{E}_{\mathrm{driv}}$ and the phase of current $\boldsymbol{J}_{\mathrm{a}}$. Due to these above, the $\Theta(\mathrm{J}_{\mathrm{a}})$ is called *modal field-current phase-mismatching factor* in this paper.

Obviously, for any single CM $\mathrm{J}_{\mathrm{a};\xi}$, there exists the following more simplified relation

$$\Theta(\mathrm{J}_{\mathrm{a};\xi}) = |\lambda_{\xi}| \tag{13}$$

and it very clearly reveals the physical meaning of $|\lambda_{\xi}|$ — the field-current phase-mismatching factor of the $\xi$-th CM itself. This implies that the CMs with smaller $|\lambda_{\xi}|$ are more desired for supplying energy from voltage driver to active element.

Conventionally, the modal significance (MS) of the $\xi$-th CM is defined as $\mathrm{MS}_{\xi} = 1/|1+j\lambda_{\xi}|$ [54], so it is obvious that

$$\mathrm{MS}_{\xi} = \frac{1}{\left|1 + j\,|\lambda_{\xi}|\right|} \tag{14}$$

because the $\lambda_{\xi}$ derived from (11) is purely real [62, Sec. 7.6]. Thus, $\mathrm{MS}_{\xi}$ is a monotonically decreasing function about $|\lambda_{\xi}|$. Then, the larger $\mathrm{MS}_{\xi}$ is, the stronger field-current phase-matching is for the $\xi$-th CM.

### *D. Numerical Verifications for WEP-CMT*

In this sub-section, the WEP-CMT proposed above is applied to a specific metallic Yagi-Uda array antenna to verify its validity. The geometry and size of the antenna are shown in Fig. 1. The MSs associated to the first four lower-order CMs calculated from the WEP-CMT are shown in Fig. 7.

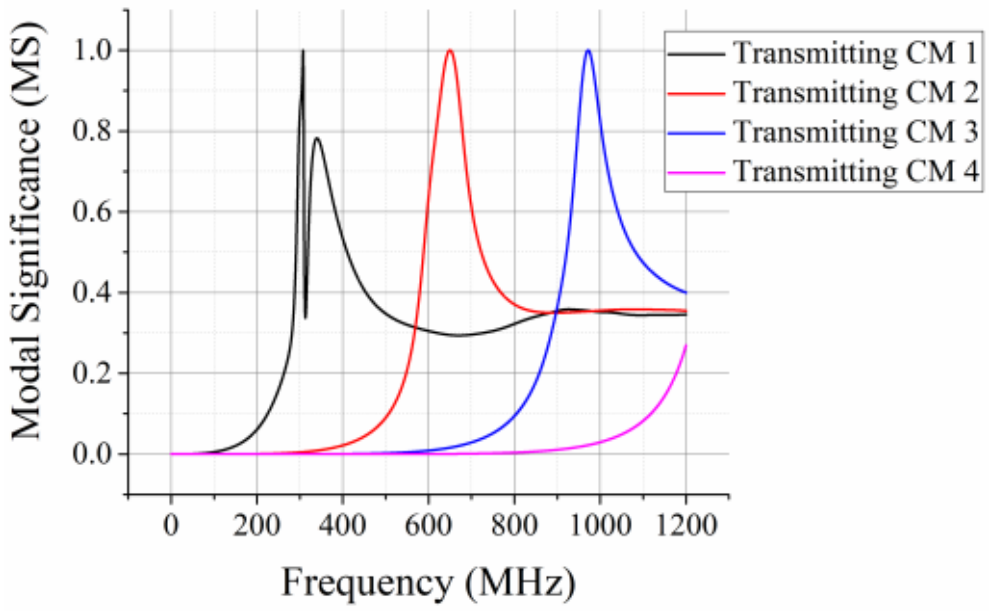

Fig. 7. MSs associated to the first four lower-order transmitting CMs calculated from the WEP-CMT proposed in this section.

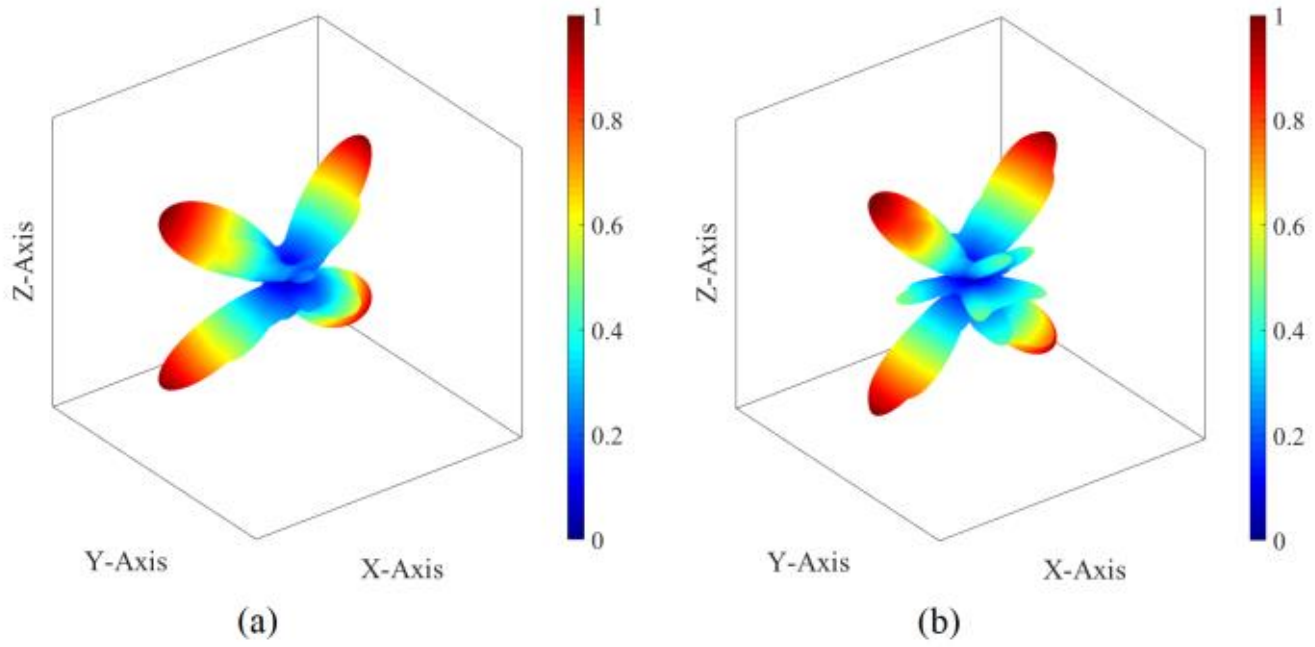

Fig. 8. Radiation patterns of the (a) resonant transmitting CM 2 working at 649.8 MHz and (b) resonant transmitting CM 3 working at 971.8 MHz.

From Fig. 7, it is easy to find out that the CM 1 is resonant at 307.3 MHz, and the resonance frequency is consistent with the one calculated from the formulation given in [6] except a 2% numerical error. The radiation pattern of the resonant CM 1 is same as the one shown in Fig. 2. In addition, the radiation patterns of the higher-order resonant CM 2 (working at 649.8 MHz) and resonant CM 3 (working at 971.8 MHz) are shown in Fig. 8. Evidently, both the resonant CM 2 and resonant CM 3 don't work at the desired end-fire state. In fact, this is just the reason why *higher resonances are available near lengths of* $\lambda$ *,* $3\lambda / 2$ *, and so forth, but are seldom used* [9, pp. 562].

In previous subsections, we have studied the difference between Yagi-Uda array *antenna* and *scatterer* from the aspects of energy source (Sec. II-A), work-energy transformation process (Sec. II-A), and CM calculation process (Sec. II-B). For further comparing the difference, we show the modal current distributions (corresponding to time-point $t = 0$) of the resonant transmitting CM (corresponding to the MS in Fig. 7 and having radiation pattern Fig. 2) and the resonant scattering CMs (corresponding to the MSs in Fig. 3 and having the far-field power-density distributions in Fig. 4) in Fig. 9 and Fig. 10.

In fact, the scattering characteristic currents in Fig. 10 appear element-self local oscillations with the variation of $t$, but the transmitting characteristic current in Fig. 9 appears a more complicated element-to-element non-local oscillation besides the element-self local oscillations. (The corresponding time-domain dynamic current figures are also uploaded to *IEEE Manuscript System* with this manuscript.) This implies

- the resonant transmitting CM originates from the cooperative work among all the dipole elements;
- a resonant scattering CM approximately corresponds to the resonant scattering state of a single dipole with some slight perturbations from the other dipoles.

This is one of the main differences between the resonant transmitting CMs and resonant scattering CMs of the 6-element linear metallic Yagi-Uda antenna discussed in this section.

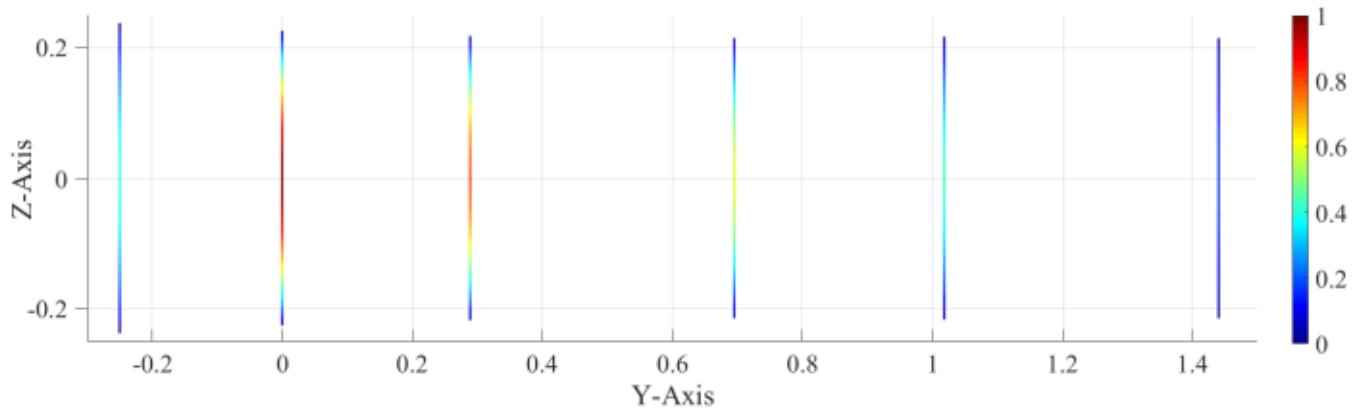

Fig. 9. Current of the resonant transmitting CM 1 (corresponding to 307.3 MHz) working at end-fire state, which has the radiation pattern shown in Fig. 2.

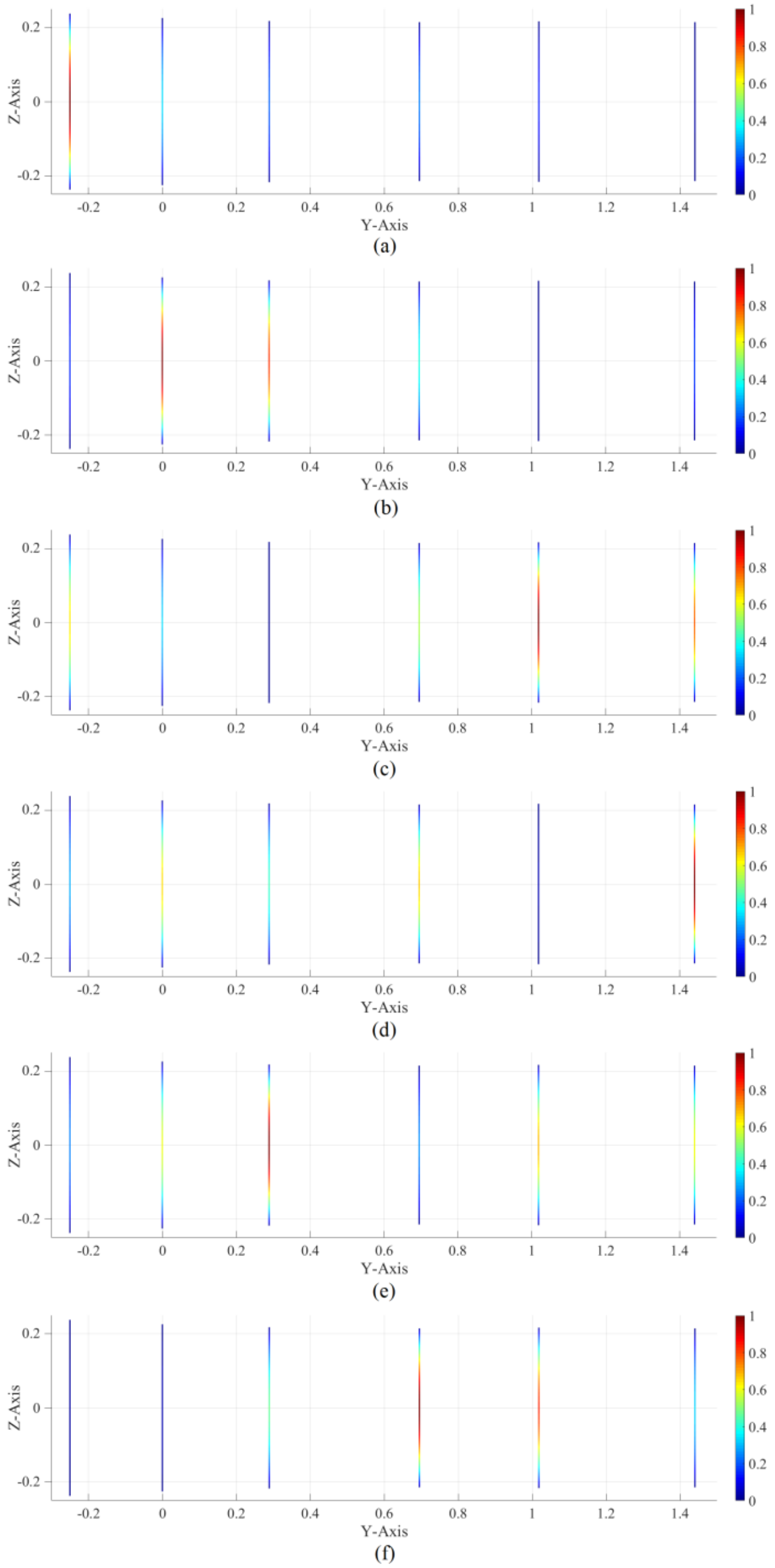

Fig. 10. Currents of the resonant scattering (a) CM 1 at 293.6 MHz, (b) CM 2 at 307.0 MHz, (c) CM 3 at 313.6 MHz, (d) CM 4 at 336.2 MHz, (e) CM 5 at 340.9 MHz, and (f) CM 6 at 345.9 MHz, which work at resonant scattering states and have the far-field power-density distributions shown in Fig. 4.

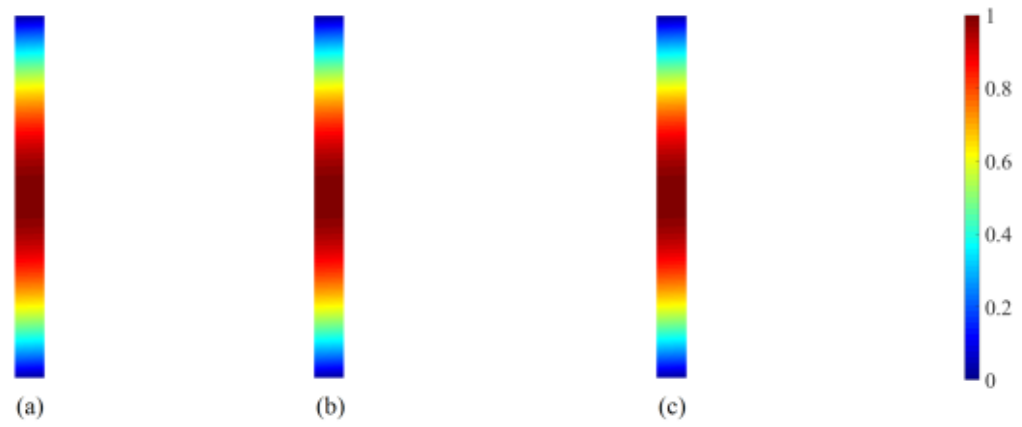

Fig. 11. Active element current distributions of the transmitting CM 1 working at (a) 307.3 MHz (resonance), (b) 312.9 MHz (local minimum of MS), and (c) 339.9 MHz (local maximum of MS but $\text{MS} \approx 0.78 < 1$).

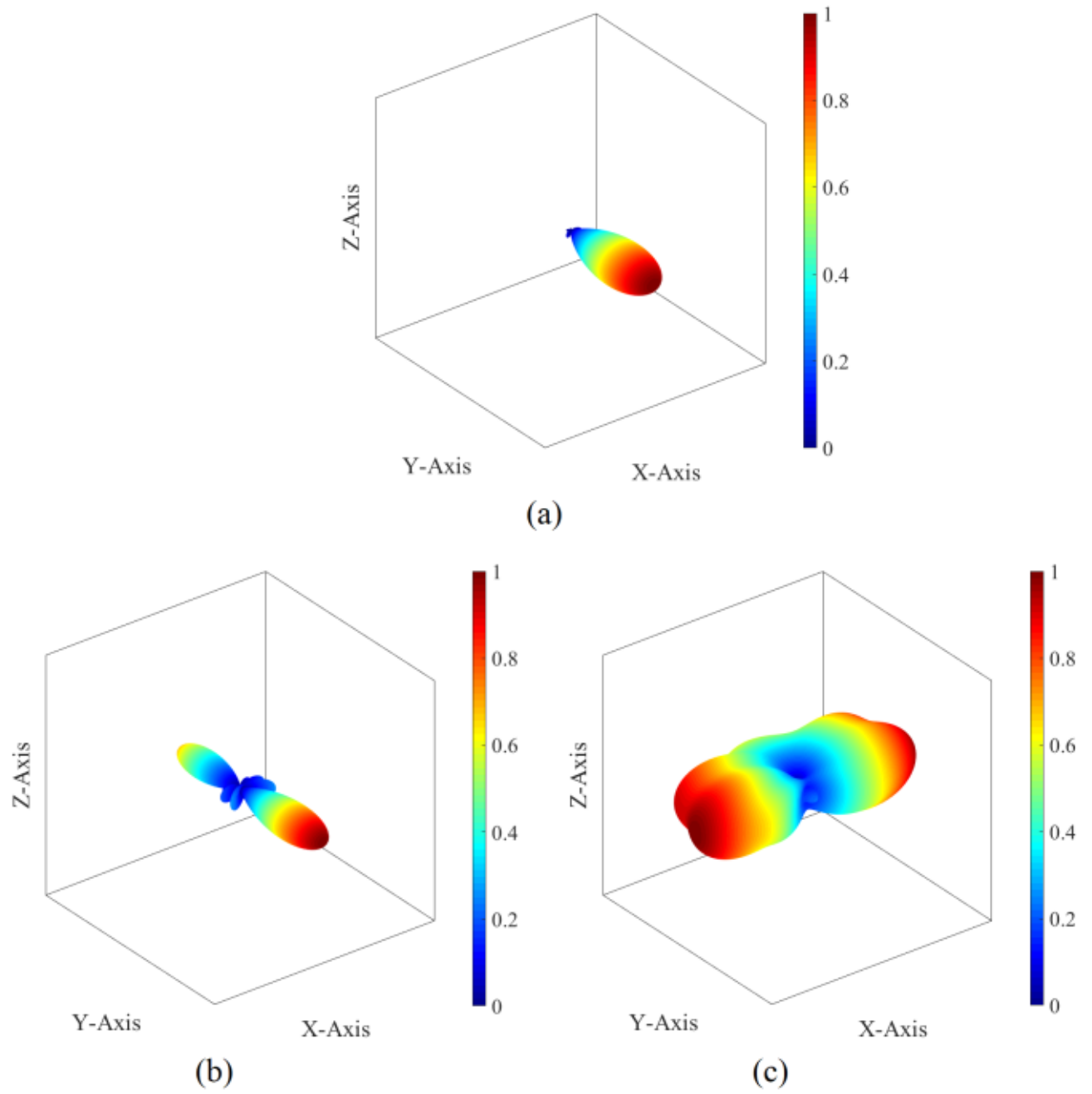

Fig. 12. Radiation patterns of the transmitting CM 1 working at (a) 307.3 MHz (resonance), (b) 312.9 MHz (local minimum of MS), and (c) 339.9 MHz (local maximum of MS but $\text{MS} \approx 0.78 < 1$).

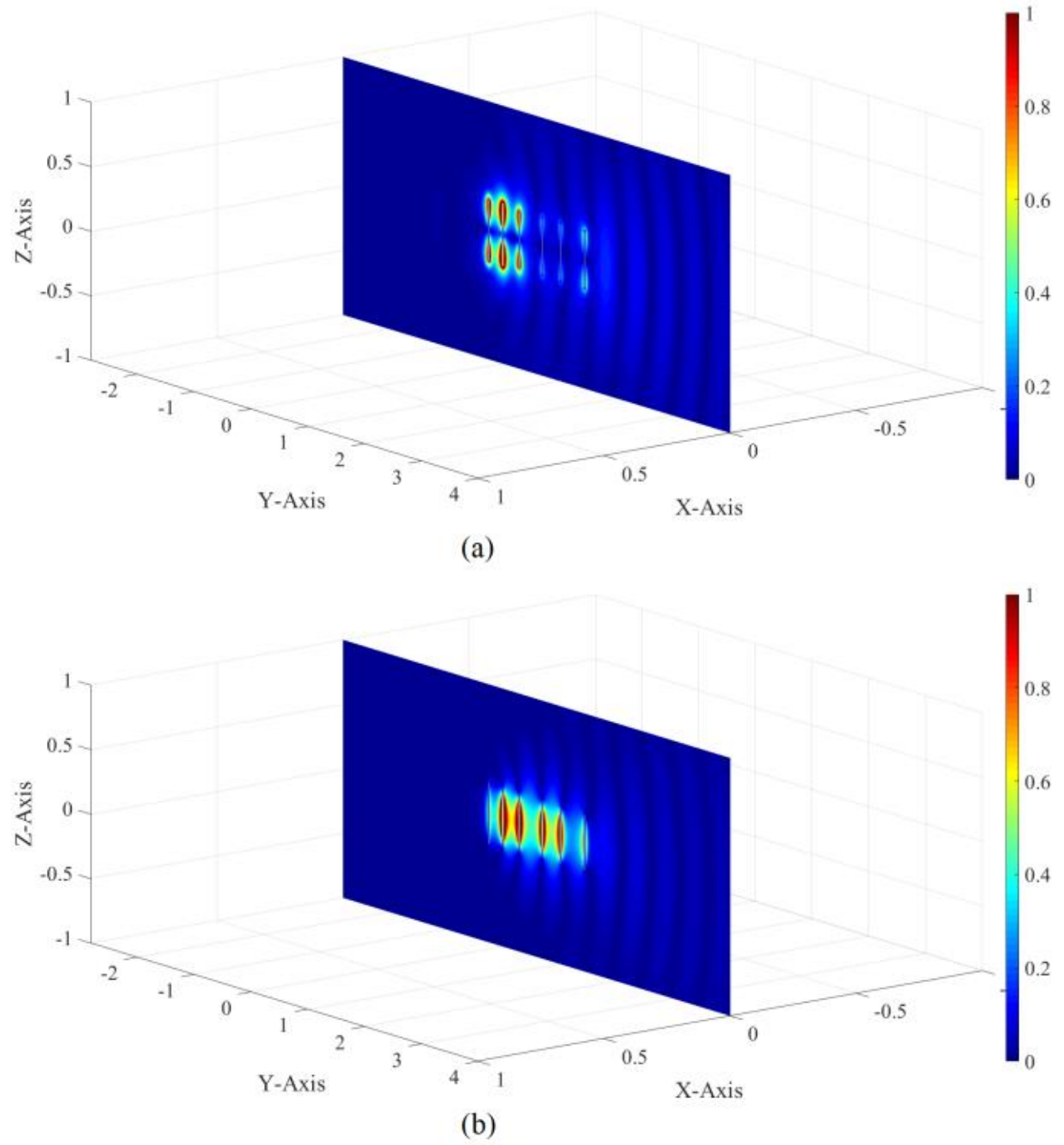

Fig. 13. Distributions of the (a) electric field and (b) magnetic field of the resonant transmitting CM 1 at end-fire state (corresponding to 307.3 MHz).

Besides the above resonant CM 1 working at 307.3 MHz, resonant CM 2 working at 649.8 MHz, and resonant CM 3 working at 971.8 MHz, there also exist another two interesting working states of the Yagi-Uda antenna, and they are the CM 1 working at 312.9 MHz (which corresponds to a local minimum of MS curve) and the CM 1 working at 339.9 MHz (which corresponds to a local maximum of MS curve but the corresponding MS is less than 0.8), and we will further investigate them as below.

For the three working states of CM 1 at 307.3 MHz, 312.9 MHz, and 339.9 MHz, their active element currents are compared in Fig. 11, and their radiation patterns are compared in Fig. 12. Evidently, the currents are similar, but only the first working state (i.e., the CM 1 at 307.3 MHz) satisfies the well-known end-fire feature of linear metallic Yagi-Uda antennas — the radiated EM power propagates along the direction from reflecting element to directing elements [7]-[10] as shown in Fig. 13, while another two working states have several non-negligible side lobes being similar to the higher-order resonant CM 2 and CM 3 shown in Fig. 8. This phenomenon will be explained below by employing Sec. II-C.

The modal significance (MS) and radiation pattern characterize CM from two different aspects.

- As physically interpreted in the previous Sec. II-C, the MS is a quantitative depiction for the matching degree between the phases of characteristic driving field and characteristic current, and the larger MS is, the stronger field-current phase-matching is. This implies that the resonant CM 1 at 307.3 MHz, resonant CM 2 at 649.8 MHz, and resonant CM 3 at 971.8 MHz have very desired field-current phase-matching character, but the CM 1 at 312.9 MHz and 339.9 MHz don't.
- As everyone knows, the radiation pattern of a working mode depends on the coherent superposition of the polarization and phase distributions of the elementary fields generated by all the elements used to constitute the antenna [7]-[10]. This implies that the elementary fields of the resonant CM 1 at 307.3 MHz tend to in-phase enhancement in the end-fire direction, and, at the same time, out-phase cancellation in the other directions. It can be similarly explained why the CM 1 at 312.9 MHz and 339.9 MHz, the CM 2 at 649.8 MHz, and the CM 3 at 971.8 MHz have several non-negligible side lobes.

Thus, the MS characterizing modal port field-current phase-matching feature and the radiation pattern characterizing modal spatial field-field coherent-superposition feature may not have a simple correspondence relation. The further studies for the relation are open to the future.

Another interesting observation from Fig. 7 is that: the MS-based maximum-minimum-maximum phenomenon appeared near the resonance frequency of the dominant CM 1 doesn't occur near the resonance frequencies of the higher-order CM 2 and CM 3. Here, we will provide a relatively clear physical picture for the phenomenon.

Qualitatively speaking, the end-fire state shown in Fig. 12(a) and Fig. 13 relies on the simultaneous satisfaction of the following two conditions:

- ◊ the reflecting element (simply denoted as RE for simplifying the following discussion) reflects EM energy along the end-fire direction;
- ◊ the directing elements (simply denoted as DEs for simplifying the following discussion) guide EM energy along the end-fire direction.

Unfortunately, although the reflecting effect of RE and the directing effect of DEs can be effectively achieved at the dominant resonant state of CM 1 (corresponding to 307.3 MHz and having end-fire radiation pattern Fig. 12(a)), the effects cannot be guaranteed in all working frequencies such as 312.9 MHz (corresponding to the axially-two-directional-fire radiation pattern Fig. 12(b)) and 339.9 MHz (corresponding to the lateral-fire radiation pattern Fig. 12(c)).

In fact, the axially-two-directional-fire feature of the CM 1 at 312.9 MHz originates from that: the RE also has some directing effects besides its well-known reflecting effect; similarly, the DEs also have some reflecting effects besides their well-known directing effect. This implies that: a part of EM energy carried by the axially-two-directional-fire CM "oscillates back and forth" between the RE and DEs, and, at the same time, the other part of EM energy is two-directionally fired along the axis of the array. (The corresponding time-domain dynamic figures are also uploaded to *IEEE Manuscript System* with this manuscript.) This *quasi* back-and-forth oscillation is very similar to the back-and-forth oscillation of the internally resonant modes of closed metallic cavities, except that the former exists some energy spills along the axial direction (as shown in the axially-two-directional-fire radiation pattern Fig. 12(b)) but the latter doesn't exist the energy spill.

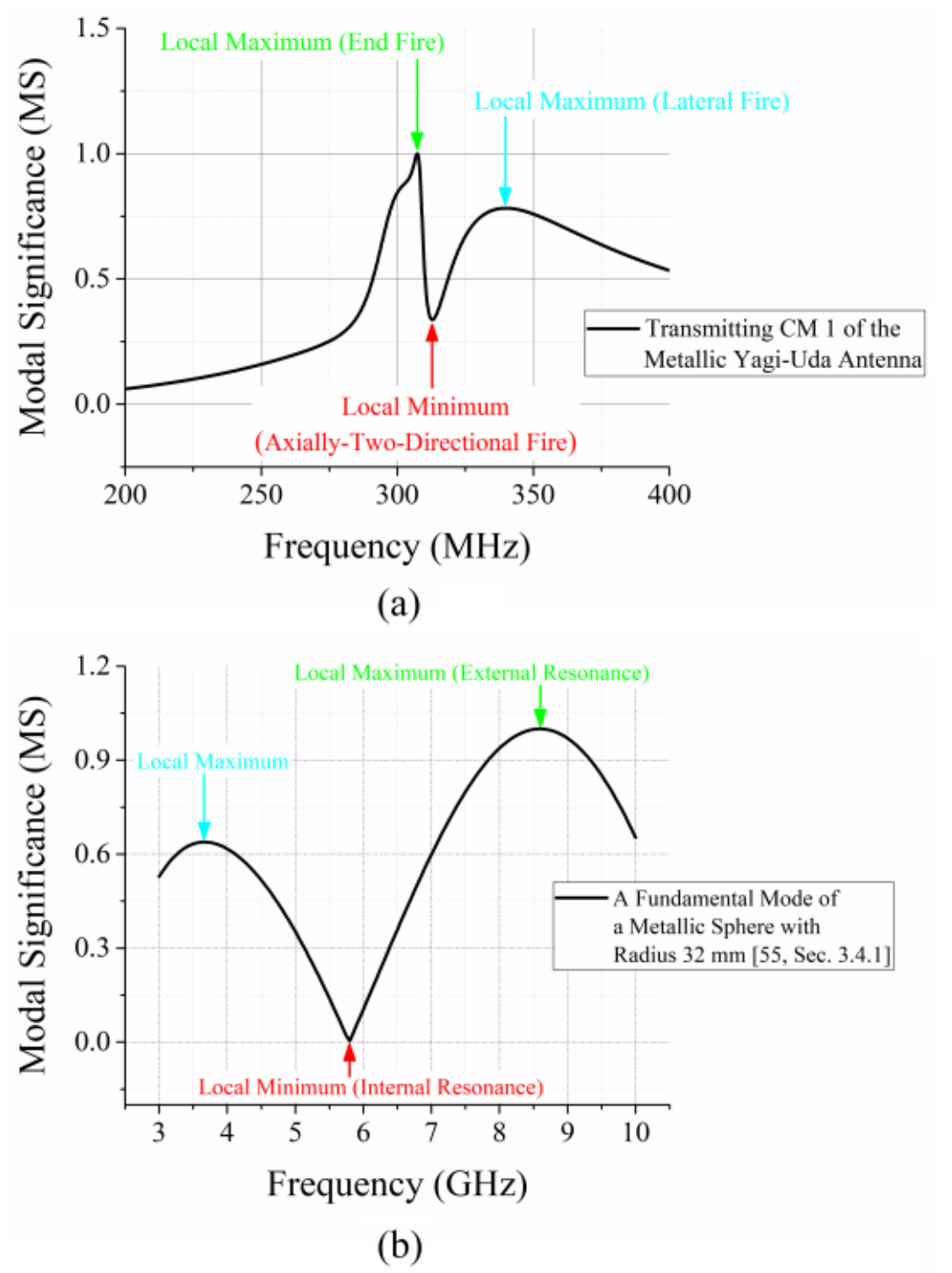

Fig. 14. MS curves of (a) the transmitting CM 1 of the metallic Yagi-Uda antenna analyzed above and (b) a fundamental mode of a metallic sphere with radius 32 mm (where the fundamental mode of metallic sphere is just the mode 2 shown in the Figure 3-33 of [55, Sec. 3.4.1]).

To confirm the above qualitative physical picture for the axially-two-directional-fire phenomenon of the CM 1 working at 312.9 MHz, we compare the MS curve of the transmitting CM 1 of the metallic Yagi-Uda antenna analyzed above and the MS curve of a fundamental mode of a metallic sphere with radius 32 mm (the modal analysis for the metallic sphere had been done in [55, Sec. 3.2.4] and [61]) in Fig. 14. Evidently, the MS-based maximum-minimum-maximum phenomenon appears in both Fig. 14(a) and Fig. 14(b). In addition, it had been proved in [55, Sec. 3.4.1] and [61] that the local minimum of the curve shown in Fig. 14(b) is corresponding to the internally resonant mode, which oscillates back and forth inside the metallic spherical cavity and doesn't radiate any EM energy.

But for the CM 2 and CM 3, their EM energies are mainly radiated in lateral directions rather than axial direction as exhibited by their radiation patterns in Figs. 8 (a) and 8(b), so the above-mentioned *quasi* back-and-forth oscillation phenomenon cannot occur at these working modes. This is just the reason why the MS-based maximum-minimum-maximum phenomenon appeared near the resonance frequency of the dominant CM 1 doesn't appear near the resonance frequencies of the higher-order CM 2 and CM 3.

### *E. WEP-CMT-Based Improvements for the Size and Bandwidth of the Metallic Yagi-Uda Antenna Discussed in Sec. II-D*

The metallic Yagi-Uda array antenna discussed above is constituted by some parallel linear dipoles, so the lateral size of the whole antenna is determined by the length of the longest linear dipole. For reducing the lateral size of whole antenna, a very natural idea is to reduce the lengths of the dipoles, but it is impracticable usually, because the change for the lengths will change the electrical performances of the antenna, such as the resonance frequency and radiation pattern of dominant mode.

In this sub-section, an alternative scheme is proposed to reduce the lateral size of the antenna, and, at the same time, the main electrical performances (such as the resonance frequency and radiation pattern of dominant mode) of the antenna can be maintained as verified by WEP-CMT-based modal analysis. The alternative scheme is to replace the linear dipoles in the original antenna with 90° folded dipoles, as shown in Fig. 15.

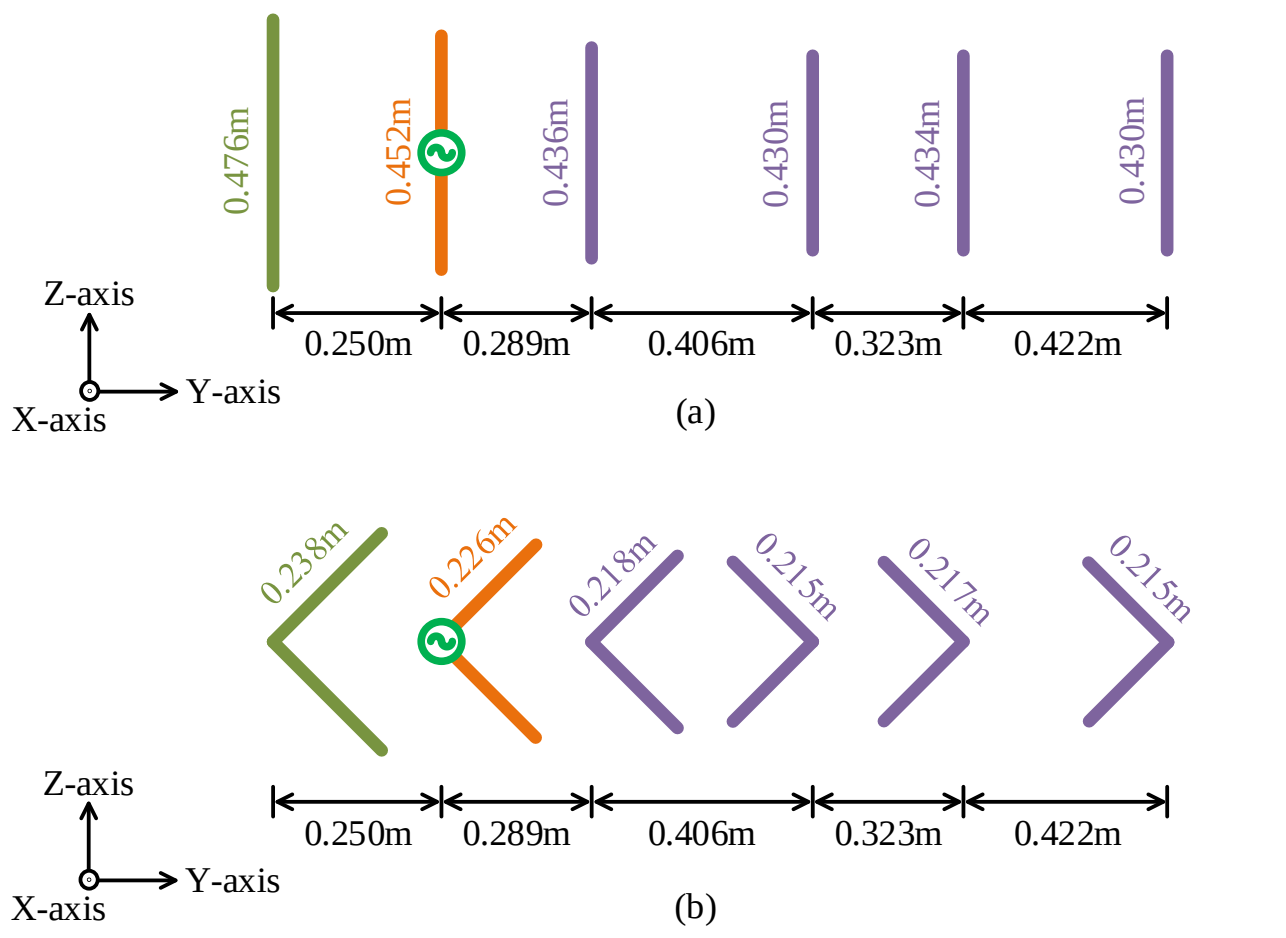

Fig. 15. Geometry and size of the Yagi-Uda array antennas constituted by (a) parallel linear dipole elements and (b) coplanar 90° folded dipole elements.

Obviously, the WEP-CMT established in this section is applicable to the folded-dipole Yagi-Uda array shown in Fig. 15(b). The WEP-CMT-based MSs of the folded-dipole array is shown in Fig. 16, and the radiation pattern of the dominant resonant mode is shown in Fig. 17. Evidently, the lateral size of the folded-dipole array is effectively reduced by 30% compared with the original linear-dipole array, and, at the same, the main electrical performances (such as the resonance frequency and radiation pattern of dominant mode) of the folded-dipole array are as satisfactory as the original linear-dipole array.

Similar to the linear-dipole array, the dominant resonant CM of the folded-dipole array also has a relatively narrow MS bandwidth, due to the existence of the MS-based local minimum near the dominant resonance frequency of CM 1. Here, a scheme is proposed to broaden the MS bandwidth of the dominant resonant CM, and the scheme is to one-by-one rotate the folded dipoles by an angle $\gamma$ (in degree) around Y-axis.

In the cases of $\gamma =$ 0°, 5°, 10°, 15°, 20°, 25°, 30°, 35°, 40°, and 45°, the WEP-CMT-based MSs associated with the dominant CM are shown in Fig. 18. Evidently, the MS bandwidth of the dominant CM can be effectively broadened as the increase of rotation angle $\gamma$.

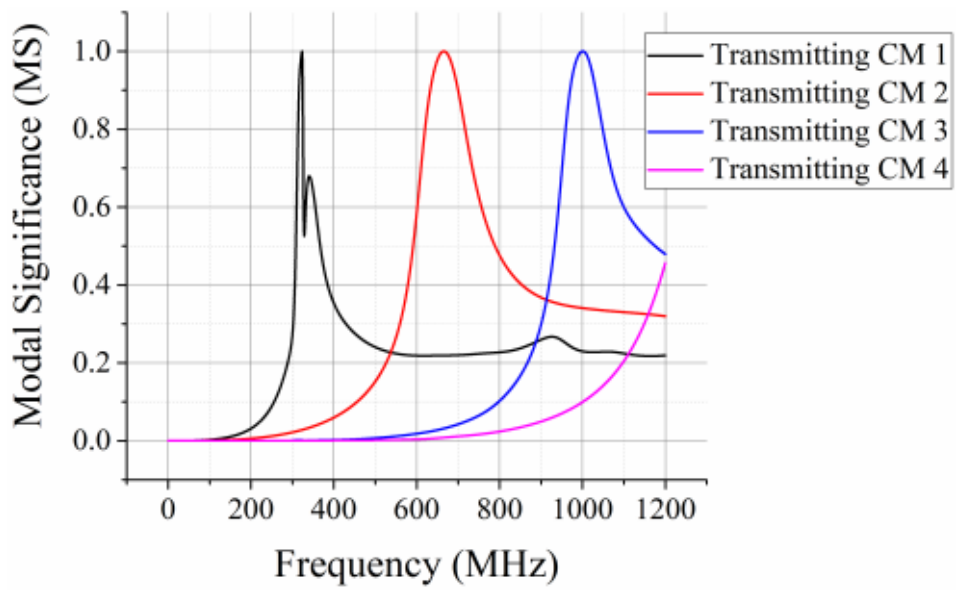

Fig. 16. MSs associated to the first four WEP-CMT-based transmitting CMs of the coplanar folded-dipole Yagi-Uda array antenna shown in Fig. 15(b).

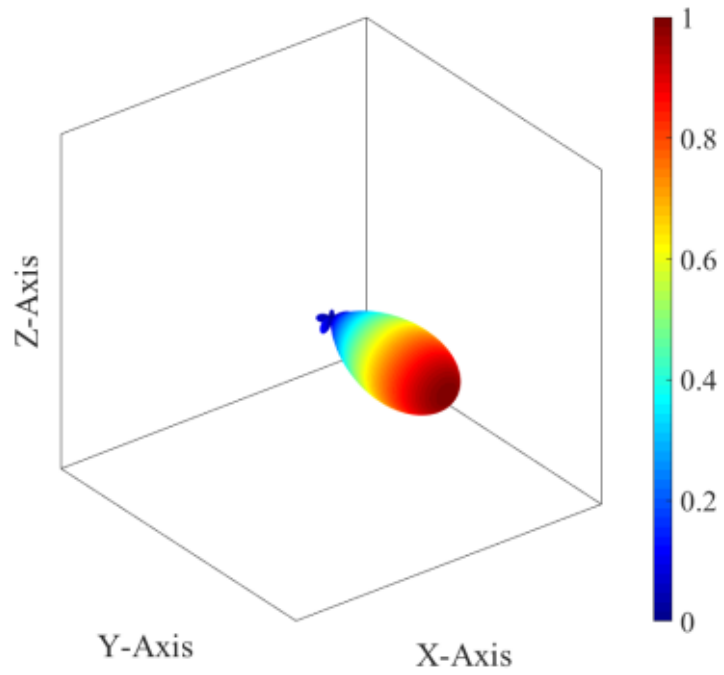

Fig. 17. Radiation pattern of the dominant resonant transmitting CM in Fig. 16.

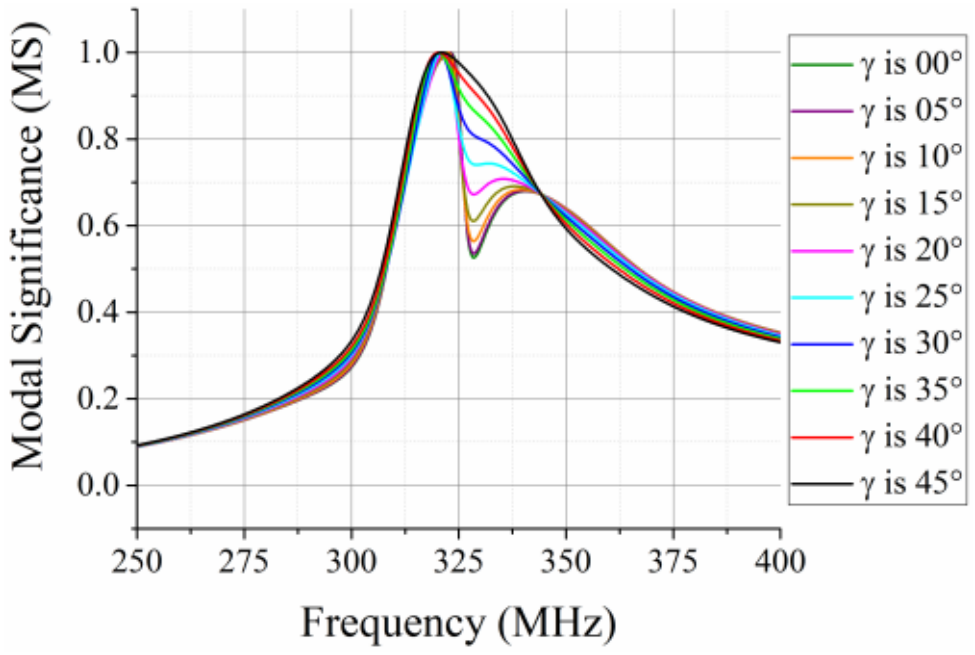

Fig. 18. MSs associated to the WEP-CMT-based dominant transmitting CM of the rotated-folded-dipole Yagi-Uda antenna with a series of rotation angles.

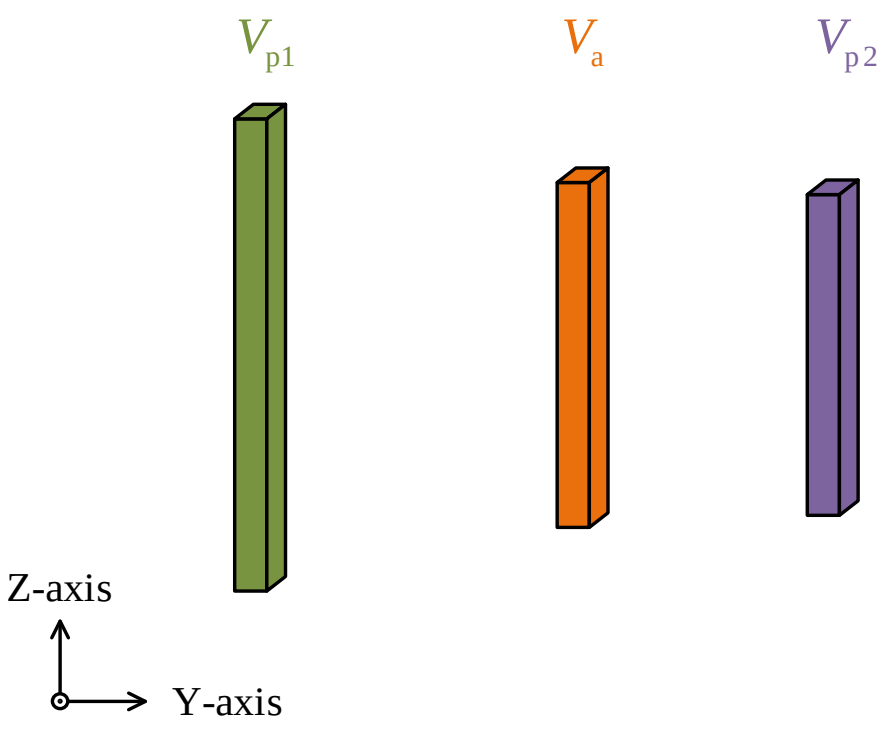

Fig. 19. Geometry of a typical 3-element linear material Yagi-Uda array antenna reported in [42]. The size and complex permittivity of the antenna are specified in Sec. III-C.

## III. WEP-CMT for Material Yagi-Uda Array Antennas

This section is devoted to generalizing the idea of Sec. II (which is for *metallic* Yagi-Uda array antennas) to *material* Yagi-Uda array antennas. A typical 3-element material Yagi-Uda antenna reported in [42] is shown in Fig. 19.

The material array has an active element $V_{\mathrm{a}}$ with boundary surface $S_{\mathrm{a}}$ and two passive/parasitic elements $V_{\mathrm{p1}}$ and $V_{\mathrm{p2}}$ with boundary surfaces $S_{\mathrm{p1}}$ and $S_{\mathrm{p2}}$ respectively. For simplifying the following discussions, the elements are restricted to being non-magnetic in this section, and their complex permittivities are denoted as $\boldsymbol{\varepsilon}_{\mathrm{a}}^{\mathrm{c}}$, $\boldsymbol{\varepsilon}_{\mathrm{p1}}^{\mathrm{c}}$, and $\boldsymbol{\varepsilon}_{\mathrm{p2}}^{\mathrm{c}}$. The purely magnetic case and magneto-dielectric case can be similarly discussed.

### A. Volume Formulation of the WEP-CMT for Material Yagi-Uda Array Antennas

If the induced volume electric currents distributing on $V_{\mathrm{a}}$, $V_{\mathrm{p1}}$, and $V_{\mathrm{p2}}$ are denoted as $\boldsymbol{j}_{\mathrm{a}}$, $\boldsymbol{j}_{\mathrm{p1}}$, and $\boldsymbol{j}_{\mathrm{p2}}$ respectively, then the corresponding DPO is as follows:

$$P_{\mathrm{driv}} = (1/2)\left\langle \boldsymbol{j}_{\mathrm{a}} , \left( j\omega\Delta\boldsymbol{\varepsilon}_{\mathrm{a}}^{\mathrm{c}} \right)^{-1} \cdot \boldsymbol{j}_{\mathrm{a}} + j\omega\mu_0 \mathcal{L}_0 \left( \boldsymbol{j}_{\mathrm{a}} + \boldsymbol{j}_{\mathrm{p1}} + \boldsymbol{j}_{\mathrm{p2}} \right) \right\rangle_{V_{\mathrm{a}}} \quad (15)$$

where $\Delta\boldsymbol{\varepsilon}_{\mathrm{a}}^{\mathrm{c}} = \boldsymbol{\varepsilon}_{\mathrm{a}}^{\mathrm{c}} - \mathbf{I}\varepsilon_0$ and $\mathbf{I}$ is the unit dyad.

Similar to deriving (10) from (7), the following

$$P_{\mathrm{driv}} = \mathrm{j}_{\mathrm{a}}^{\dagger} \cdot \mathbb{P}_{\mathrm{driv}} \cdot \mathrm{j}_{\mathrm{a}} \quad (16)$$

with only independent current $\mathrm{j}_{\mathrm{a}}$ can be derived from (15), and a detailed derivation process is given in App. D. Here, $\mathrm{j}_{\mathrm{a}}$ is the basis function expansion coefficient vector of $\boldsymbol{j}_{\mathrm{a}}$.

Employing the $\mathbb{P}_{\mathrm{driv}}$ in (16), the CMs of the material Yagi-Uda antenna can be calculated from the equation like (11).

The scheme provided in this sub-section is based on volume currents, and an alternative surface-current-based scheme will be given in the following sub-section.

### B. Surface Formulation of the WEP-CMT for Material Yagi-Uda Array Antennas

If the equivalent surface currents distributing on $S_{\mathrm{a}}$, $S_{\mathrm{p1}}$, and $S_{\mathrm{p2}}$ are denoted as $(\boldsymbol{J}_{\mathrm{a}}, \boldsymbol{M}_{\mathrm{a}})$, $(\boldsymbol{J}_{\mathrm{p1}}, \boldsymbol{M}_{\mathrm{p1}})$, and $(\boldsymbol{J}_{\mathrm{p2}}, \boldsymbol{M}_{\mathrm{p2}})$ respectively, then the volume-current version (15) of DPO can be alternatively written as the following surface-current version

$$P_{\rm driv} = -(1/2)\left\langle \boldsymbol{J}_{\rm a}, \mathcal{E}_0\left(\boldsymbol{J}_{\rm a}+\boldsymbol{J}_{\rm p1}+\boldsymbol{J}_{\rm p2}, \boldsymbol{M}_{\rm a}+\boldsymbol{M}_{\rm p1}+\boldsymbol{M}_{\rm p2}\right)\right\rangle_{S_{\rm a}^-} - (1/2)\left\langle \boldsymbol{M}_{\rm a}, \mathcal{H}_0\left(\boldsymbol{J}_{\rm a}+\boldsymbol{J}_{\rm p1}+\boldsymbol{J}_{\rm p2}, \boldsymbol{M}_{\rm a}+\boldsymbol{M}_{\rm p1}+\boldsymbol{M}_{\rm p2}\right)\right\rangle_{S_{\rm a}^-} \quad (17)$$

in which operator $\mathcal{E}_0$ is as $\mathcal{E}_0(\boldsymbol{J},\boldsymbol{M}) = -j\omega\mu_0\mathcal{L}_0(\boldsymbol{J}) - \mathcal{K}_0(\boldsymbol{M})$ and operator $\mathcal{H}_0$ is as $\mathcal{H}_0(\boldsymbol{J},\boldsymbol{M}) = \mathcal{K}_0(\boldsymbol{J}) - j\omega\varepsilon_0\mathcal{L}_0(\boldsymbol{M})$, where $\mathcal{L}_0$ is the same as the one used previously and $\mathcal{K}_0$ is defined as that $\mathcal{K}_0(\boldsymbol{X}) = \nabla\times\int_\Omega G_0(\boldsymbol{r},\boldsymbol{r}')\boldsymbol{X}(\boldsymbol{r}')d\Omega'$. In addition, the equivalent surface currents are defined by employing the inner normal directions of the boundaries of the array elements.

Similar to deriving (10) from (7), the following

$$P_{\rm driv} = \mathbb{M}_{\rm a}^\dagger \cdot \mathbb{P}_{\rm driv} \cdot \mathbb{M}_{\rm a} \quad (18)$$

with only independent current $\mathbb{M}_{\rm a}$ can be derived from (17), and a detailed derivation process is given in App. E. Here, $\mathbb{M}_{\rm a}$ is the basis function expansion coefficient vector of $\boldsymbol{M}_{\rm a}$. Theoretically, the independent current can also be selected as the $\mathbb{J}_{\rm a}$, which is the basis function expansion coefficient vector of $\boldsymbol{J}_{\rm a}$. However, the numerical performances of the two selections are different, and it is more desirable to select $\mathbb{M}_{\rm a}$ as independent current because the array elements are non-magnetic, and a similar explanation focusing on scattering objects can be found in [55, Sec. 6.2] and [58].

Employing the $\mathbb{P}_{\rm driv}$ in (18), the CMs of the material Yagi-Uda antenna can be calculated from the equation like (11).

### C. Numerical Verifications

To verify the validities of the above volume and surface formulations of the WEP-CMT for material Yagi-Uda antennas, the comparisons of the WEP-CMT-based numerical results with some published simulation and measurement data are provided in this sub-section. The specific material Yagi-Uda antenna (with the geometry shown in Fig. 19) analyzed here is the same as the one reported in [42], and its all elements are with $4.0\text{ mm}\times4.0\text{ mm}$ cross section, and its elements $V_{\rm a}$, $V_{\rm p1}$, and $V_{\rm p2}$ have lengths 46.35 mm, 77.6 mm, and 44.4 mm respectively, and the distance (side-to-side) between $V_{\rm a}$ and $V_{\rm p1}$ is 23.0 mm, and the distance (side-to-side) between $V_{\rm a}$ and $V_{\rm p2}$ is 10.7 mm. The complex permittivities of the array elements are that $\boldsymbol{\varepsilon}_{\rm a}^{\rm c} = \boldsymbol{\varepsilon}_{\rm p1}^{\rm c} = \boldsymbol{\varepsilon}_{\rm p2}^{\rm c} = \mathbf{I}34\varepsilon_0$.

The characteristic value (in decibel) of the dominant CM calculated from WEP-CMT and the modal $S_{11}$ parameter (in decibel) obtained from simulation and measurement published in [42] are shown in Fig. 20 simultaneously. From the figure, it is easy to find out that the WEP-CMT-based resonance frequency is basically consistent with the data reported in [42], and the slight discrepancy is mainly originated from ignoring the feeding structure.

For the resonant state of the WEP-CMT-based CM shown in Fig. 20, its radiation pattern is shown in Fig. 21, and its electric and magnetic field distributions are shown in Fig. 22. Evidently, Fig. 21 and Fig. 22 satisfy the well-known end-fire feature of linear Yagi-Uda arrays — the radiated power propagates along the direction from reflecting element to directing elements.

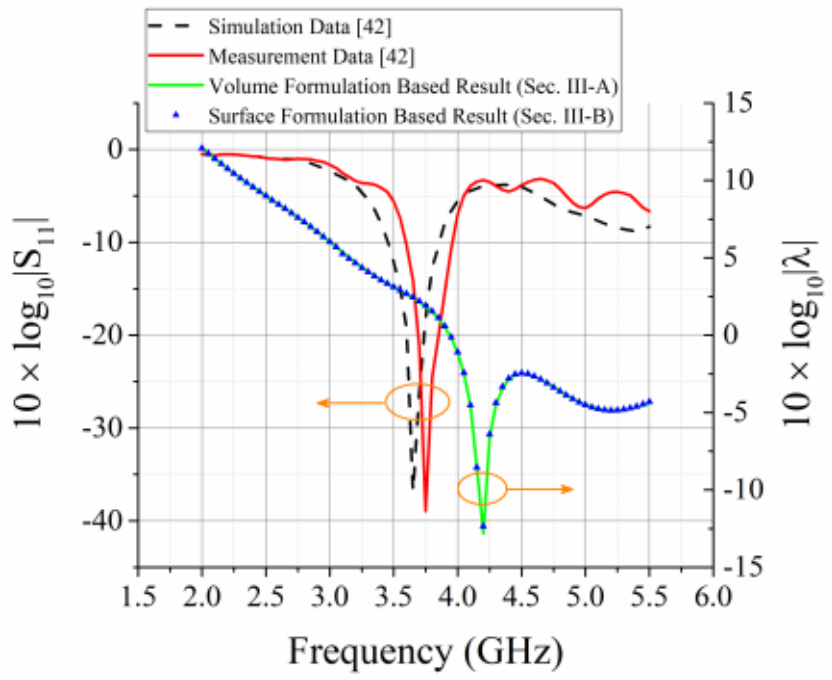

Fig. 20. Modal parameters of the dominant transmitting mode of the material Yagi-Uda array antenna reported in [42].

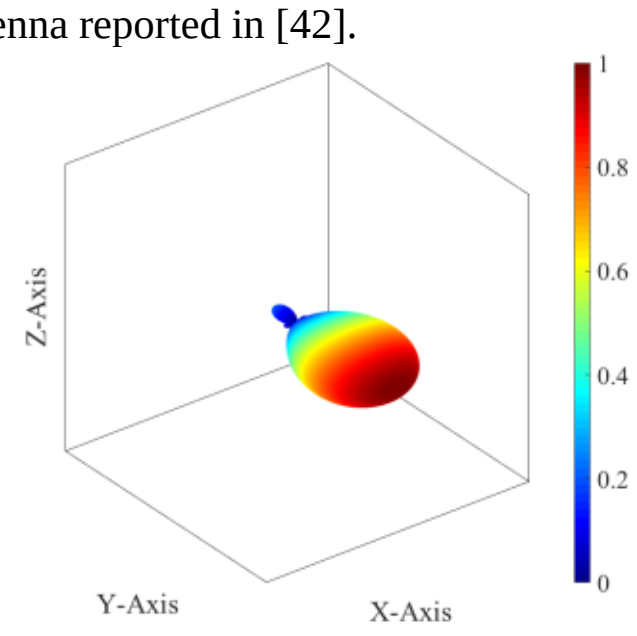

Fig. 21. Radiation pattern of the WEP-CMT-based resonant transmitting CM.

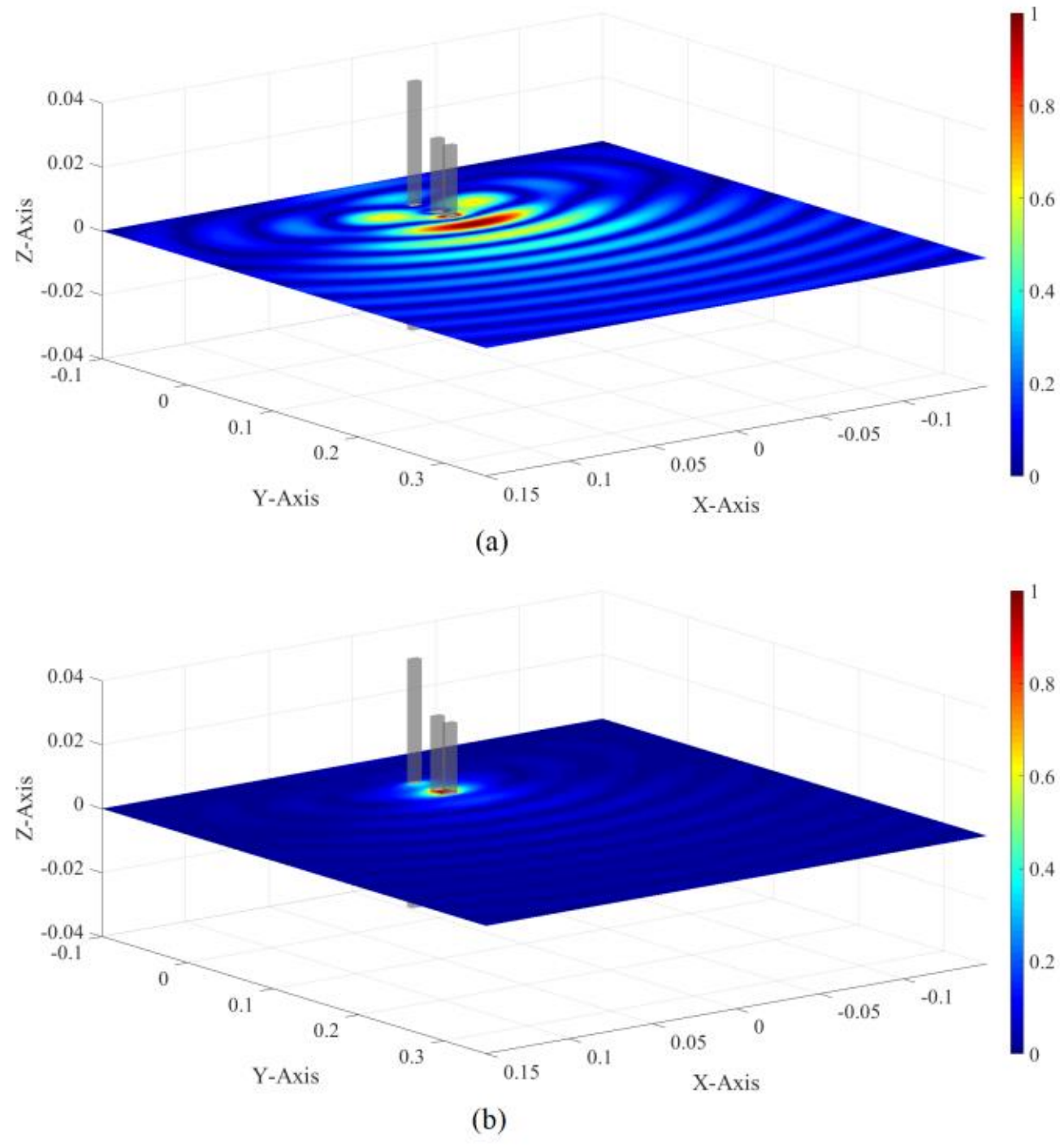

Fig. 22. Distributions of the (a) electric field and (b) magnetic field of the resonant state of the WEP-CMT-based transmitting CM shown in Fig. 20.

## IV. WEP-CMT for Composite Yagi-Uda Array Antennas

This section further generalizes the results given in Secs. II and III to the *metal-material composite* Yagi-Uda array antenna shown in Fig. 23, which was reported in [23]. The antenna is constituted by metallic active patch $S_{\rm a}$, metallic passive patches $S_{\rm p}$, metallic ground plane $V_{\rm g}$, and material substrate $V_{\rm s}$. The substrate $V_{\rm s}$ is restricted to being non-magnetic for simplifying the discussions, and its complex permittivity is $\boldsymbol{\varepsilon}_{\rm s}^{\rm c}$.

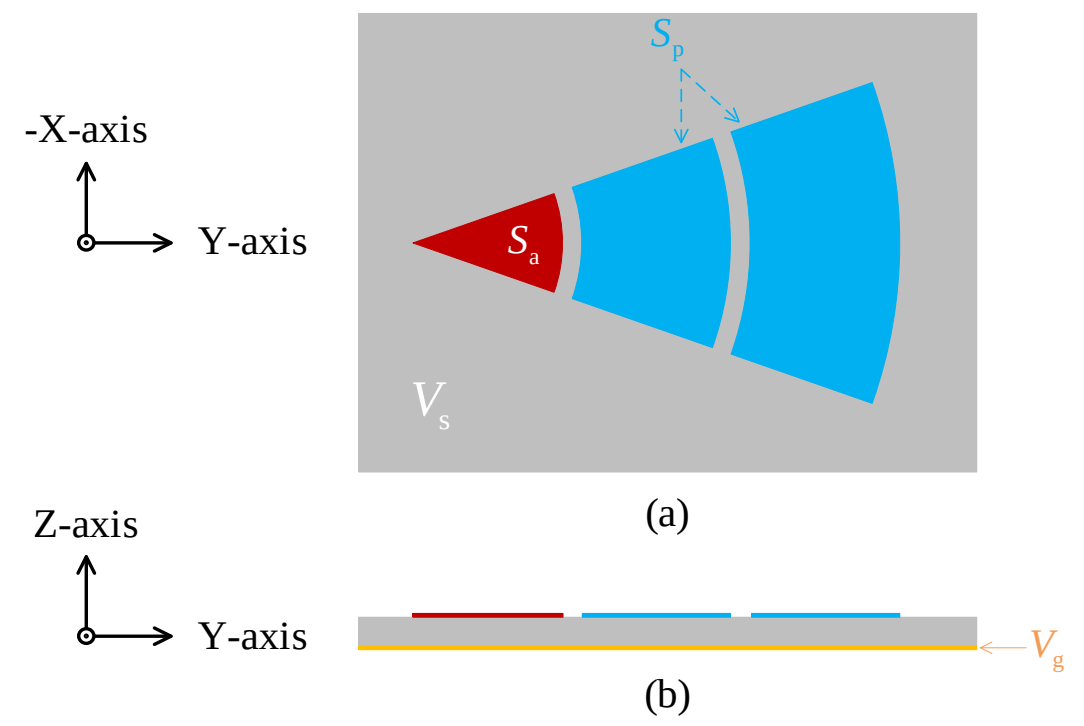

Fig. 23. Geometry of a 3-element composite Yagi-Uda array antenna reported in [23]. (a) Top view; (b) lateral view. The detailed size is given in [23], and the complex permittivity of the substrate is $\boldsymbol{\varepsilon}_s^c = \mathbf{I}2.55\varepsilon_0$ .

## A. *Volume-Surface Formulation of the WEP-CMT for Composite Yagi-Uda Array Antennas*

If the induced surface electric currents on $S_a$ , $S_p$ , and the boundary of $V_g$ are denoted as $\boldsymbol{J}_a$ , $\boldsymbol{J}_p$ , and $\boldsymbol{J}_g$ respectively, and the induced volume electric current on $V_s$ is denoted as $\boldsymbol{j}_s$ , then the corresponding DPO is as follows:

$$P_{driv} = (1/2)\left\langle \boldsymbol{J}_a , j\omega\mu_0\mathcal{L}_0\left(\boldsymbol{J}_a + \boldsymbol{J}_p + \boldsymbol{J}_g \oplus \boldsymbol{j}_s\right)\right\rangle_{S_a} \tag{19}$$

Here, " $\oplus$ " is to emphasize the difference between the dimensions of surface current $\boldsymbol{J}_a + \boldsymbol{J}_p + \boldsymbol{J}_g$ and volume current $\boldsymbol{j}_s$ .

Similar to deriving (10) from (7), the following

$$P_{driv} = \mathrm{J}_a^\dagger \cdot \mathrm{P}_{driv} \cdot \mathrm{J}_a \tag{20}$$

with only independent current $\mathrm{J}_a$ can be derived from (19), and a detailed derivation process is given in App. F. Here, $\mathrm{J}_a$ is the basis function expansion coefficient vector of $\boldsymbol{J}_a$ .

The formula provided in this sub-section is based on volume-surface currents, and an alternative surface-current-based formula will be given in the following sub-section.

## B. *Surface Formulation of the WEP-CMT for Composite Yagi-Uda Array Antennas*

For the convenience of this sub-section, the interface between $V_g$ and free space is denoted as $S_{gf}$ , and the interface between $V_s$ and $S_a$ is denoted as $S_{sa}$ , and the interface between $V_s$ and $S_p$ is denoted as $S_{sp}$ , and the interface between $V_s$ and free space is denoted as $S_{sf}$ .

If the induced surface electric current on $S_{gf}$ is denoted as $\boldsymbol{J}_{gf}$ , and the equivalent surface electric currents on $S_{sa}$ , $S_{sp}$ , and $S_{sf}$ are denoted as $\boldsymbol{J}_{sa}$ , $\boldsymbol{J}_{sp}$ , and $\boldsymbol{J}_{sf}$ respectively, and the equivalent surface magnetic current on $S_{sf}$ is denoted as $\boldsymbol{M}_{sf}$ , then the volume-surface-current version (19) of DPO can be alternatively written as the following surface-current version

$$P_{driv} = \frac{1}{2}\left\langle \boldsymbol{J}_a , -\mathcal{E}_0\left(\boldsymbol{J}_a + \boldsymbol{J}_p + \boldsymbol{J}_{gf} - \boldsymbol{J}_{sa} - \boldsymbol{J}_{sp} - \boldsymbol{J}_{sf}, -\boldsymbol{M}_{sf}\right)\right\rangle_{S_a} \tag{21}$$

where the equivalent currents are defined by using the inner normal direction of the boundary of the material substrate.

Similar to deriving (10) from (7), the following matrix-formed DPO

$$P_{driv} = \mathrm{J}_a^\dagger \cdot \mathrm{P}_{driv} \cdot \mathrm{J}_a \tag{22}$$

with only independent current $\mathrm{J}_a$ can be derived from (21), and a detailed derivation process is given in App. G.

Employing the $\mathrm{P}_{driv}$ used in (20) or (22), the CMs of the composite Yagi-Uda antenna can be calculated from the characteristic equation like (11).

## C. *Numerical Verifications*

This sub-section applies the WEP-CMT to the composite Yagi-Uda antenna reported in [23]. The geometry of the antenna is shown in Fig. 23; the size of the antenna is described in [23]; the complex permittivity of the dielectric substrate of the composite antenna is $\boldsymbol{\varepsilon}_s^c = \mathbf{I}2.55\varepsilon_0$ .

The characteristic value (in decibel) of the dominant CM calculated from WEP-CMT and the modal $S_{11}$ parameter (in decibel) obtained from simulation published in [23] are shown in Fig. 24 simultaneously. The figure implies that the WEP-CMT-based resonance frequencies are basically consistent with the data reported in [23], and the slight discrepancy is mainly originated from ignoring the feeding structure. The multi-resonance phenomenon shown in Fig. 24 is because of exciting different CMs in different frequency bands.

The radiation pattern of the CM working at dominant resonance frequency 5.12 GHz is shown in Fig. 25.

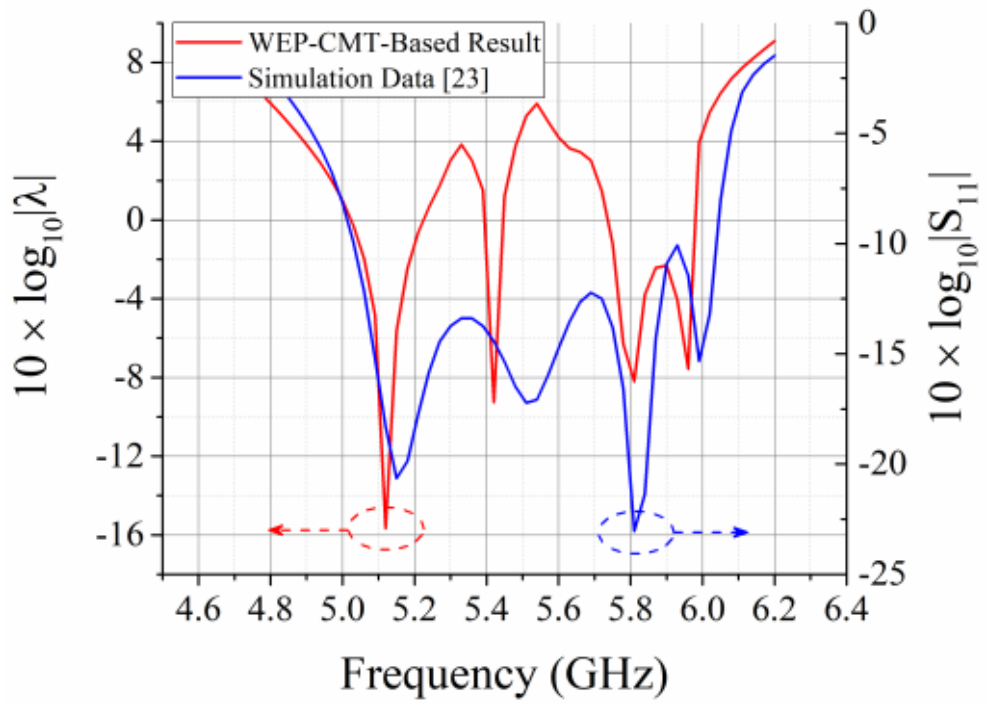

Fig. 24. Modal parameters of the working mode of the composite Yagi-Uda array antenna reported in [23].

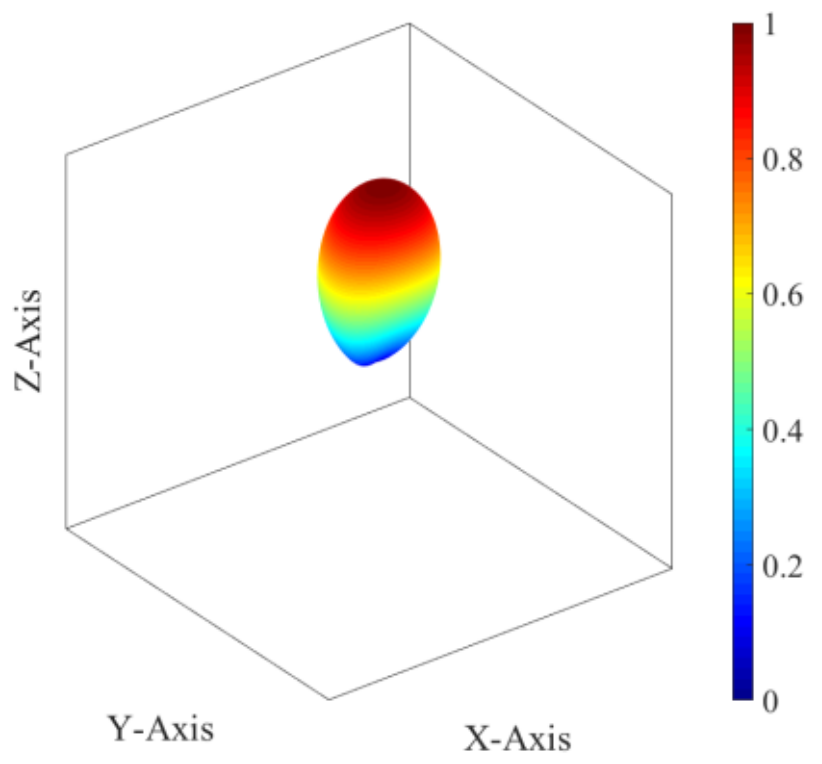

Fig. 25. Radiation pattern of the dominant resonant state (at 5.12 GHz) of the WEP-CMT-based CM shown in Fig. 24.

## V. Conclusions

Similar to the conclusions of our previous studies for scattering objects, the WEP is also a quantitative depiction for the work-energy transformation process of Yagi-Uda array antennas, and the driving power is just the source to sustain a steady work-energy transformation. Employing the scatterer-oriented and antenna-oriented WEPs and driving powers, it is found out that the working mechanisms of scattering objects (such as the Yagi-Uda *array scatterer* shown in Fig. 5(a)) and Yagi-Uda *array antennas* are different from each other as reflected in the aspects of energy source (Sec. II-A) and work-energy transformation process (Sec. II-A). The different working mechanisms expose that the conventional CMT for scattering objects cannot be simply applied to Yagi-Uda array antennas.

Under WEP framework, the conventional WEP-based CMT for scattering objects is generalized to metallic Yagi-Uda array antennas, by using some necessary modifications. By orthogonalizing DPO, the generalized WEP-CMT can construct a set of energy-decoupled *transmitting* CMs for pre-selected objective antenna. For effectively constructing CMs, the dependent currents contained in DPO must be eliminated (specifically, the currents on parasitic elements must be expressed in terms of the functions of the current on active element) before solving characteristic equation, and this is one of the main differences between the CM calculation processes for metallic Yagi-Uda array antennas and metallic scattering objects. By comparing the CM calculation process (Sec. II-B) and modal current distribution (Sec. II-D) of the Yagi-Uda arrays viewed as transmitting antenna and as scattering object, the conclusion "the conventional CMT for scattering objects cannot be simply applied to Yagi-Uda array antennas" is further reconfirmed.

In addition, the generalized WEP-CMT for metallic Yagi-Uda array antennas is also further generalized to material and composite Yagi-Uda array antennas successfully.

By employing the WEP-based modal decomposition and the field-current interaction expression of driving power, it is clearly revealed that the characteristic values / modal significances associated with the WEP-CMT-based CMs of metallic Yagi-Uda antennas are the quantitative depiction for the phase-mismatching / phase-matching degree between characteristic driving fields and characteristic currents. Moreover, the WEP-based physical interpretation for the characteristic values and modal significances of metallic Yagi-Uda antennas is also applicable to the material and composite Yagi-Uda antennas.

The validity of the generalized WEP-CMT is verified by applying the WEP-CMT to the various Yagi-Uda antennas and comparing the WEP-CMT-based results with the published analytical, simulation, and measurement data. For a classical 6-element metallic antenna, a WEP-CMT-based modal analysis implies that the dominant resonant CM is end-fire but the other higher-order resonant CMs are usually not, and then it is clearly explained why the higher-order resonant modes are seldom used; the physical picture depicting the MS-based maximum-minimum-maximum phenomenon appeared near the dominant resonance frequency is provided; it is also explained why the maximum-minimum-maximum phenomenon doesn't appear around the resonance frequencies of higher-order CMs.

## Appendices

Some detailed formulations related to this paper are provided in the following appendices.

### A. Rigorous Derivation for Work-Energy Principle (1)

Based on the working mechanism of the metallic Yagi-Uda antennas discussed in Sec. II-A, we can obtain the following relations

$$\begin{aligned}(1/2)\langle \boldsymbol{J}_{\rm a}, \boldsymbol{E}_{\rm driv}\rangle_{S_{\rm a}} &= -(1/2)\langle \boldsymbol{J}_{\rm a}, \boldsymbol{E}\rangle_{S_{\rm a}} \\ &= -(1/2)\langle \boldsymbol{J}_{\rm a}, \boldsymbol{E}\rangle_{S_{\rm a}} - (1/2)\langle \boldsymbol{J}_{\rm p}, \boldsymbol{E}\rangle_{S_{\rm p}} \\ &= -(1/2)\langle \boldsymbol{J}_{\rm a}+\boldsymbol{J}_{\rm p}, \boldsymbol{E}\rangle_{S_{\rm a}\cup S_{\rm p}} \\ &= (1/2)\oiint_{S_\infty}(\boldsymbol{E}\times\boldsymbol{H}^\dagger)\cdot d\boldsymbol{S} \\ &\quad + j2\omega\left[(1/4)\langle \boldsymbol{H},\boldsymbol{B}\rangle_{\mathbb{E}^3} - (1/4)\langle \boldsymbol{D},\boldsymbol{E}\rangle_{\mathbb{E}^3}\right]\end{aligned} \tag{23}$$

In (23), the first equality is due to that the tangential component of $\boldsymbol{E}_{\rm driv}+\boldsymbol{E}$ is zero on $S_{\rm a}$; the second equality is due to that the tangential component of $\boldsymbol{E}$ is zero on $S_{\rm p}$; the third equality is obvious due to the linear property of inner product; the last equality originates from Poynting's theorem because $\boldsymbol{J}_{\rm a}+\boldsymbol{J}_{\rm p}$ is just the source to generate $\boldsymbol{E}$, where $\boldsymbol{E}$ is the summation of $\boldsymbol{E}_{\rm a}$ (generated by $\boldsymbol{J}_{\rm a}$) and $\boldsymbol{E}_{\rm p}$ (generated by $\boldsymbol{J}_{\rm p}$) as explained in Sec. II-A.

### B. Detailed Formulations Related to Sec. II-B

In (8), the elements of sub-matrix $\mathbb{P}_{\rm aa}$ are calculated as that $[\mathbb{P}_{\rm aa}]_{\xi\zeta} = (1/2) < \boldsymbol{b}_{{\rm a};\xi}, j\omega\mu_0\mathcal{L}_0(\boldsymbol{b}_{{\rm a};\zeta}) >_{S_{\rm a}}$, and the elements of sub-matrix $\mathbb{P}_{\rm ap}$ are calculated as that $[\mathbb{P}_{\rm ap}]_{\xi\zeta} = (1/2) < \boldsymbol{b}_{{\rm a};\xi}, j\omega\mu_0\mathcal{L}_0(\boldsymbol{b}_{{\rm p};\zeta}) >_{S_{\rm a}}$.

The $\mathbb{T}$ used in (9) is $\mathbb{T} = -\mathbb{Z}_{\rm pp}^{-1}\cdot\mathbb{Z}_{\rm pa}$. The elements of matrix $\mathbb{Z}_{\rm pp}$ are calculated as that $[\mathbb{Z}_{\rm pp}]_{\xi\zeta} =< \boldsymbol{b}_{{\rm p};\xi}, -j\omega\mu_0\mathcal{L}_0(\boldsymbol{b}_{{\rm p};\zeta}) >_{S_{\rm p}}$, and the elements of matrix $\mathbb{Z}_{\rm pa}$ are calculated as that $[\mathbb{Z}_{\rm pa}]_{\xi\zeta} =< \boldsymbol{b}_{{\rm p};\xi}, -j\omega\mu_0\mathcal{L}_0(\boldsymbol{b}_{{\rm a};\zeta}) >_{S_{\rm p}}$.

The $\mathbb{P}_{\rm driv}$ used in (10) is as follows:

$$\mathbb{P}_{\rm driv} = \begin{bmatrix}\mathbb{P}_{\rm aa} & \mathbb{P}_{\rm ap}\end{bmatrix}\cdot\begin{bmatrix}\mathbb{I} \\ \mathbb{T}\end{bmatrix} \tag{24}$$

where $\mathbb{I}$ is an unit matrix with proper order.

### C. Rigorous Derivations for the Radiated Power Orthogonality (5) and Reactive Power Orthogonality (6) Satisfied by Metallic Yagi-Uda Array Antennas

Repeating the process given in [55, Sec. 3.2.3], it is easy to prove that the CMs derived from (11) satisfy the following orthogonality relations

$$\delta_{\xi\zeta} = \mathrm{J}_{{\rm a};\xi}^\dagger\cdot\mathbb{P}_{\rm driv}^+\cdot\mathrm{J}_{{\rm a};\zeta} \tag{25}$$

$$\lambda_\xi\,\delta_{\xi\zeta} = \mathrm{J}_{{\rm a};\xi}^\dagger\cdot\mathbb{P}_{\rm driv}^-\cdot\mathrm{J}_{{\rm a};\zeta} \tag{26}$$

Substituting $\mathbb{P}_{\rm driv}^+ = [\mathbb{P}_{\rm driv}+\mathbb{P}_{\rm driv}^\dagger]/2$ into (25), it is immediate to have that

$$\begin{aligned}\delta_{\xi\zeta} &= (1/2)\mathbb{J}_{a;\xi}^{\dagger}\cdot\mathbb{P}_{driv}\cdot\mathbb{J}_{a;\zeta}+(1/2)\mathbb{J}_{a;\xi}^{\dagger}\cdot\mathbb{P}_{driv}^{\dagger}\cdot\mathbb{J}_{a;\zeta}\\ &= (1/2)\mathbb{J}_{a;\xi}^{\dagger}\cdot\mathbb{P}_{driv}\cdot\mathbb{J}_{a;\zeta}+(1/2)\left[\mathbb{J}_{a;\zeta}^{\dagger}\cdot\mathbb{P}_{driv}\cdot\mathbb{J}_{a;\xi}\right]^{\dagger}\\ &= \frac{1}{2}\mathbb{J}_{a;\xi}^{\dagger}\cdot\left[\mathbb{P}_{aa}\ \mathbb{P}_{ap}\right]\cdot\begin{bmatrix}\mathbb{J}_{a;\zeta}\\ \mathbb{J}_{p;\zeta}\end{bmatrix}+\frac{1}{2}\left\{\mathbb{J}_{a;\zeta}^{\dagger}\cdot\left[\mathbb{P}_{aa}\ \mathbb{P}_{ap}\right]\cdot\begin{bmatrix}\mathbb{J}_{a;\xi}\\ \mathbb{J}_{p;\xi}\end{bmatrix}\right\}^{\dagger}\\ &= (1/2)\left[\mathbb{J}_{a;\xi}^{\dagger}\cdot\mathbb{P}_{aa}\cdot\mathbb{J}_{a;\zeta}+\mathbb{J}_{a;\xi}^{\dagger}\cdot\mathbb{P}_{ap}\cdot\mathbb{J}_{p;\zeta}\right]\\ &\quad+(1/2)\left[\mathbb{J}_{a;\zeta}^{\dagger}\cdot\mathbb{P}_{aa}\cdot\mathbb{J}_{a;\xi}+\mathbb{J}_{a;\zeta}^{\dagger}\cdot\mathbb{P}_{ap}\cdot\mathbb{J}_{p;\xi}\right]^{\dagger}\end{aligned} \tag{27}$$

In (27), the second equality is based on the †-operation-based *reverse-order law* of matrix algebra [62, Sec. 0.2.5]; the third equality is due to (9) and (24), where $\mathbb{J}_{p;\zeta}=\mathbb{T}\cdot\mathbb{J}_{a;\zeta}$ and $\mathbb{J}_{p;\xi}=\mathbb{T}\cdot\mathbb{J}_{a;\xi}$; the last equality is obvious.

Matrix element $[\mathbb{P}_{aa}]_{\xi\zeta}=(1/2)<\boldsymbol{b}_{a;\xi},j\omega\mu_0\mathcal{L}_0(\boldsymbol{b}_{a;\zeta})>_{S_a}$ is just the inner product between basis current $\boldsymbol{b}_{a;\xi}$ and the electric field generated by basis current $\boldsymbol{b}_{a;\zeta}$ (with a coefficient $-1/2$), so the term $\mathbb{J}_{a;\xi}^{\dagger}\cdot\mathbb{P}_{aa}\cdot\mathbb{J}_{a;\zeta}$ in (27) is just the inner product between characteristic current $\boldsymbol{J}_{a;\xi}$ and the electric field generated by characteristic current $\boldsymbol{J}_{a;\zeta}$ (with a coefficient $-1/2$). The terms $\mathbb{J}_{a;\xi}^{\dagger}\cdot\mathbb{P}_{ap}\cdot\mathbb{J}_{p;\zeta}$, $\mathbb{J}_{a;\zeta}^{\dagger}\cdot\mathbb{P}_{aa}\cdot\mathbb{J}_{a;\xi}$, and $\mathbb{J}_{a;\zeta}^{\dagger}\cdot\mathbb{P}_{ap}\cdot\mathbb{J}_{p;\xi}$ in (27) can be similarly interpreted.

Based on the above physical interpretations for the terms involved in (27) and employing the fact that the tangential $\boldsymbol{E}$ is zero on $S_p$, the (27) can be equivalently written as the following alternative forms

$$\begin{aligned}\delta_{\xi\zeta} &= (1/2)\left\{-(1/2)\langle\boldsymbol{J}_{\xi},\boldsymbol{E}_{\zeta}\rangle_{S_a\cup S_p}-(1/2)\langle\boldsymbol{J}_{\zeta},\boldsymbol{E}_{\xi}\rangle_{S_a\cup S_p}^{\dagger}\right\}\\ &= (1/2)\Big\{(1/2)\oiint_{S_\infty}\left(\boldsymbol{E}_{\zeta}\times\boldsymbol{H}_{\xi}^{\dagger}\right)\cdot d\boldsymbol{S}\\ &\qquad+j2\omega\left[(\mu_0/4)\langle\boldsymbol{H}_{\xi},\boldsymbol{H}_{\zeta}\rangle_{\mathbb{E}^3}-(\varepsilon_0/4)\langle\boldsymbol{E}_{\xi},\boldsymbol{E}_{\zeta}\rangle_{\mathbb{E}^3}\right]\Big\}\\ &\quad+(1/2)\Big\{(1/2)\oiint_{S_\infty}\left(\boldsymbol{E}_{\xi}\times\boldsymbol{H}_{\zeta}^{\dagger}\right)\cdot d\boldsymbol{S}\\ &\qquad+j2\omega\left[(\mu_0/4)\langle\boldsymbol{H}_{\zeta},\boldsymbol{H}_{\xi}\rangle_{\mathbb{E}^3}-(\varepsilon_0/4)\langle\boldsymbol{E}_{\zeta},\boldsymbol{E}_{\xi}\rangle_{\mathbb{E}^3}\right]\Big\}^{\dagger}\\ &= (1/2)\Big\{(1/2\eta_0)\langle\boldsymbol{E}_{\xi},\boldsymbol{E}_{\zeta}\rangle_{S_\infty}\\ &\qquad+j2\omega\left[(\mu_0/4)\langle\boldsymbol{H}_{\xi},\boldsymbol{H}_{\zeta}\rangle_{\mathbb{E}^3}-(\varepsilon_0/4)\langle\boldsymbol{E}_{\xi},\boldsymbol{E}_{\zeta}\rangle_{\mathbb{E}^3}\right]\Big\}\\ &\quad+(1/2)\Big\{(1/2\eta_0)\langle\boldsymbol{E}_{\xi},\boldsymbol{E}_{\zeta}\rangle_{S_\infty}\\ &\qquad-j2\omega\left[(\mu_0/4)\langle\boldsymbol{H}_{\xi},\boldsymbol{H}_{\zeta}\rangle_{\mathbb{E}^3}-(\varepsilon_0/4)\langle\boldsymbol{E}_{\xi},\boldsymbol{E}_{\zeta}\rangle_{\mathbb{E}^3}\right]\Big\}\\ &= (1/2\eta_0)\langle\boldsymbol{E}_{\xi},\boldsymbol{E}_{\zeta}\rangle_{S_\infty}\end{aligned} \tag{28}$$

where $\boldsymbol{J}_{\xi}=\boldsymbol{J}_{a;\xi}+\boldsymbol{J}_{p;\xi}$ and $\boldsymbol{J}_{\zeta}=\boldsymbol{J}_{a;\zeta}+\boldsymbol{J}_{p;\zeta}$ for simplifying the symbolic system of this App. C, and $\boldsymbol{E}_{\xi}$ and $\boldsymbol{E}_{\zeta}$ are the fields generated by $\boldsymbol{J}_{\xi}$ and $\boldsymbol{J}_{\zeta}$ respectively. In (28), the derivation for the second equality is similar to deriving the last equality in (23); the third equality is based on Sommerfeld's radiation condition [57, Sec. 3-5] and the basic operation law about complex conjugation [63, Sec. 1.4]; the last equality is obvious.

The derivation for the reactive power orthogonality (6) from (26) only needs a simple repetition of the above derivation for the radiated power orthogonality (5) from (25), so it will not be exhibited here.

## D. Detailed Formulations Related to Sec. III-A

By expanding the currents in (15) in terms of some proper basis functions, the integral form (15) of DPO can be discretized into the following matrix form

$$P_{driv}=\mathbb{j}_a^{\dagger}\cdot\left[\mathbb{P}_{aa}\quad\mathbb{P}_{ap1}\quad\mathbb{P}_{ap2}\right]\cdot\begin{bmatrix}\mathbb{j}_a\\ \mathbb{j}_{p1}\\ \mathbb{j}_{p2}\end{bmatrix} \tag{29}$$

where the elements of sub-matrix $\mathbb{P}_{aa}$ are calculated as that $[\mathbb{P}_{aa}]_{\xi\zeta}=(1/2)<\boldsymbol{b}_{a;\xi},(j\omega\Delta\boldsymbol{\varepsilon}_a^c)^{-1}\cdot\boldsymbol{b}_{a;\zeta}+j\omega\mu_0\mathcal{L}_0(\boldsymbol{b}_{a;\zeta})>_{V_a}$, and the elements of sub-matrix $\mathbb{P}_{ap1/ap2}$ are calculated as that $[\mathbb{P}_{ap1/ap2}]_{\xi\zeta}=(1/2)<\boldsymbol{b}_{a;\xi},j\omega\mu_0\mathcal{L}_0(\boldsymbol{b}_{p1/p2;\zeta})>_{V_a}$.

Because of volume equivalence principle, there exist the following integral equations

$$\boldsymbol{j}_{p1}=j\omega\Delta\boldsymbol{\varepsilon}_{p1}^c\cdot\left[-j\omega\mu_0\mathcal{L}_0\left(\boldsymbol{j}_a+\boldsymbol{j}_{p1}+\boldsymbol{j}_{p2}\right)\right]\ \text{on}\ V_{p1} \tag{30}$$

$$\boldsymbol{j}_{p2}=j\omega\Delta\boldsymbol{\varepsilon}_{p2}^c\cdot\left[-j\omega\mu_0\mathcal{L}_0\left(\boldsymbol{j}_a+\boldsymbol{j}_{p1}+\boldsymbol{j}_{p2}\right)\right]\ \text{on}\ V_{p2} \tag{31}$$

Applying the method of moments (MoM) to (30)-(31), the integral equations are immediately discretized into matrix equations. By solving the matrix equations, the following transformation

$$\begin{bmatrix}\mathbb{j}_{p1}\\ \mathbb{j}_{p2}\end{bmatrix}=\underbrace{\begin{bmatrix}\mathbb{Z}_{p1p1}&\mathbb{Z}_{p1p2}\\ \mathbb{Z}_{p2p1}&\mathbb{Z}_{p2p2}\end{bmatrix}\cdot\begin{bmatrix}\mathbb{Z}_{p1a}\\ \mathbb{Z}_{p2a}\end{bmatrix}}_{\mathbb{T}}\cdot\mathbb{j}_a \tag{32}$$

can be easily obtained, where the elements of $\mathbb{Z}_{p1p1/p2p2}$ are $<\boldsymbol{b}_{p1/p2;\xi},(j\omega\Delta\boldsymbol{\varepsilon}_{p1/p2}^c)^{-1}\cdot\boldsymbol{b}_{p1/p2;\zeta}+j\omega\mu_0\mathcal{L}_0(\boldsymbol{b}_{p1/p2;\zeta})>_{V_{p1}/V_{p2}}$, and the elements of sub-matrix $\mathbb{Z}_{p1p2/p2p1}$ are calculated as $<\boldsymbol{b}_{p1/p2;\xi},j\omega\mu_0\mathcal{L}_0(\boldsymbol{b}_{p2/p1;\zeta})>_{V_{p1}/V_{p2}}$, and the elements of $\mathbb{Z}_{p1a/p2a}$ are $<\boldsymbol{b}_{p1/p2;\xi},-j\omega\mu_0\mathcal{L}_0(\boldsymbol{b}_{a;\zeta})>_{V_{p1}/V_{p2}}$.

Substituting (32) into (29), we immediately have (16), where

$$\mathbb{P}_{driv}=\left[\mathbb{P}_{aa}\quad\mathbb{P}_{ap1}\quad\mathbb{P}_{ap2}\right]\cdot\begin{bmatrix}\mathbb{I}\\ \mathbb{T}\end{bmatrix} \tag{33}$$

## E. Detailed Formulations Related to Sec. III-B

The integral form (17) of DPO can be discretized into the following matrix form

$$P_{driv}=\begin{bmatrix}\mathbb{J}_a\\ \mathbb{M}_a\end{bmatrix}^{\dagger}\cdot\begin{bmatrix}\mathbb{P}_{aa}^{JJ}&\mathbb{P}_{ap1}^{JJ}&\mathbb{P}_{ap2}^{JJ}&\mathbb{P}_{aa}^{JM}&\mathbb{P}_{ap1}^{JM}&\mathbb{P}_{ap2}^{JM}\\ \mathbb{P}_{aa}^{MJ}&\mathbb{P}_{ap1}^{MJ}&\mathbb{P}_{ap2}^{MJ}&\mathbb{P}_{aa}^{MM}&\mathbb{P}_{ap1}^{MM}&\mathbb{P}_{ap2}^{MM}\end{bmatrix}\cdot\begin{bmatrix}\mathbb{J}_a\\ \mathbb{J}_{p1}\\ \mathbb{J}_{p2}\\ \mathbb{M}_a\\ \mathbb{M}_{p1}\\ \mathbb{M}_{p2}\end{bmatrix} \tag{34}$$

by expanding the currents in terms of some proper basis functions. The formulations for calculating the elements of the various sub-matrices $\mathbb{P}$ in (34) are trivial, and they are not explicitly given here.

For the currents involved in (17), there exist the following integral equations

$$\left[\mathcal{E}_{\rm a}\left(\boldsymbol{J}_{\rm a},\boldsymbol{M}_{\rm a}\right)\right]_{S_{\rm a}^-}^{\rm tan}=\boldsymbol{n}_{\rm a}^-\times\boldsymbol{M}_{\rm a} \tag{35}$$

$$\begin{aligned}&\left[\mathcal{E}_{\rm p1}\left(\boldsymbol{J}_{\rm p1},\boldsymbol{M}_{\rm p1}\right)\right]_{S_{\rm p1}^-}^{\rm tan}\\&=-\left[\mathcal{E}_0\left(\boldsymbol{J}_{\rm a}+\boldsymbol{J}_{\rm p1}+\boldsymbol{J}_{\rm p2},\boldsymbol{M}_{\rm a}+\boldsymbol{M}_{\rm p1}+\boldsymbol{M}_{\rm p2}\right)\right]_{S_{\rm p1}^+}^{\rm tan}\end{aligned} \tag{36}$$

$$\begin{aligned}&\left[\mathcal{H}_{\rm p1}\left(\boldsymbol{J}_{\rm p1},\boldsymbol{M}_{\rm p1}\right)\right]_{S_{\rm p1}^-}^{\rm tan}\\&=-\left[\mathcal{H}_0\left(\boldsymbol{J}_{\rm a}+\boldsymbol{J}_{\rm p1}+\boldsymbol{J}_{\rm p2},\boldsymbol{M}_{\rm a}+\boldsymbol{M}_{\rm p1}+\boldsymbol{M}_{\rm p2}\right)\right]_{S_{\rm p1}^+}^{\rm tan}\end{aligned} \tag{37}$$

$$\begin{aligned}&\left[\mathcal{E}_{\rm p2}\left(\boldsymbol{J}_{\rm p2},\boldsymbol{M}_{\rm p2}\right)\right]_{S_{\rm p2}^-}^{\rm tan}\\&=-\left[\mathcal{E}_0\left(\boldsymbol{J}_{\rm a}+\boldsymbol{J}_{\rm p1}+\boldsymbol{J}_{\rm p2},\boldsymbol{M}_{\rm a}+\boldsymbol{M}_{\rm p1}+\boldsymbol{M}_{\rm p2}\right)\right]_{S_{\rm p2}^+}^{\rm tan}\end{aligned} \tag{38}$$

$$\begin{aligned}&\left[\mathcal{H}_{\rm p2}\left(\boldsymbol{J}_{\rm p2},\boldsymbol{M}_{\rm p2}\right)\right]_{S_{\rm p2}^-}^{\rm tan}\\&=-\left[\mathcal{H}_0\left(\boldsymbol{J}_{\rm a}+\boldsymbol{J}_{\rm p1}+\boldsymbol{J}_{\rm p2},\boldsymbol{M}_{\rm a}+\boldsymbol{M}_{\rm p1}+\boldsymbol{M}_{\rm p2}\right)\right]_{S_{\rm p2}^+}^{\rm tan}\end{aligned} \tag{39}$$

where operator $\mathcal{E}_{\rm a}$ is with parameters $(\boldsymbol{\varepsilon}_{\rm s}^{\rm c},\mu_0)$, and operators $\mathcal{E}_{\rm p1/p2}$ and $\mathcal{H}_{\rm p1/p2}$ are with parameters $(\boldsymbol{\varepsilon}_{\rm p1/p2}^{\rm c},\mu_0)$, and $S_{\rm a/p1/p2}^-$ is the inner surface of $S_{\rm a/p1/p2}$, and $S_{\rm p1/p2}^+$ is the outer surface of $S_{\rm p1/p2}$.

Similar to deriving (32) from (30) and (31), the following

$$\begin{bmatrix}\mathrm{J}_{\rm a}\\\mathrm{J}_{\rm p1}\\\mathrm{J}_{\rm p2}\\\mathrm{M}_{\rm a}\\\mathrm{M}_{\rm p1}\\\mathrm{M}_{\rm p2}\end{bmatrix}=\mathrm{T}\cdot\mathrm{M}_{\rm a} \tag{40}$$

can be derived from (35)-(39), and then we have the following sub-transformations

$$\mathrm{J}_{\rm a}=\mathrm{T}_1\cdot\mathrm{M}_{\rm a},\ \text{and}\ \begin{bmatrix}\mathrm{J}_{\rm p1}\\\mathrm{J}_{\rm p2}\end{bmatrix}=\mathrm{T}_2\cdot\mathrm{M}_{\rm a},\ \text{and}\ \begin{bmatrix}\mathrm{M}_{\rm p1}\\\mathrm{M}_{\rm p2}\end{bmatrix}=\mathrm{T}_3\cdot\mathrm{M}_{\rm a} \tag{41}$$

Substituting (41) into (34), we immediately have (18), where the $\mathrm{P}_{\rm driv}$ is as follows:

$$\mathrm{P}_{\rm driv}=\begin{bmatrix}\mathrm{T}_1\\\mathrm{I}\end{bmatrix}^\dagger\cdot\begin{bmatrix}\mathrm{P}_{\rm aa}^{\rm JJ}&\mathrm{P}_{\rm ap1}^{\rm JJ}&\mathrm{P}_{\rm ap2}^{\rm JJ}&\mathrm{P}_{\rm aa}^{\rm JM}&\mathrm{P}_{\rm ap1}^{\rm JM}&\mathrm{P}_{\rm ap2}^{\rm JM}\\\mathrm{P}_{\rm aa}^{\rm MJ}&\mathrm{P}_{\rm ap1}^{\rm MJ}&\mathrm{P}_{\rm ap2}^{\rm MJ}&\mathrm{P}_{\rm aa}^{\rm MM}&\mathrm{P}_{\rm ap1}^{\rm MM}&\mathrm{P}_{\rm ap2}^{\rm MM}\end{bmatrix}\cdot\begin{bmatrix}\mathrm{T}_1\\\mathrm{T}_2\\\mathrm{I}\\\mathrm{T}_3\end{bmatrix} \tag{42}$$

### *F. Detailed Formulations Related to Sec. IV-A*

Here, we only provide the integral equations used to establish the transformation from the independent current to the dependent currents involved in (19) as follows:

$$0=\left[-j\omega\mu_0\mathcal{L}_0\left(\boldsymbol{J}_{\rm a}+\boldsymbol{J}_{\rm p}+\boldsymbol{J}_{\rm g}+\boldsymbol{j}_{\rm s}\right)\right]^{\rm tan}\quad \text{on}\ S_{\rm p}\cup S_{\rm g} \tag{43}$$

$$\boldsymbol{j}_{\rm s}=j\omega\Delta\boldsymbol{\varepsilon}_{\rm s}^{\rm c}\cdot\left[-j\omega\mu_0\mathcal{L}_0\left(\boldsymbol{J}_{\rm a}+\boldsymbol{J}_{\rm p}+\boldsymbol{J}_{\rm g}+\boldsymbol{j}_{\rm s}\right)\right]\ \text{on}\ V_{\rm s} \tag{44}$$

and the other detailed formulations related to Sec. IV-A are similar to the ones used in Apps. B, D, and E.

### *G. Detailed Formulations Related to Sec. IV-B*

Here, we only provide the integral equations used to establish the transformation from the independent current to the dependent currents involved in (21) as follows:

$$\begin{aligned}&\left[\mathcal{E}_0\left(\boldsymbol{J}_{\rm a}+\boldsymbol{J}_{\rm p}+\boldsymbol{J}_{\rm gf}-\boldsymbol{J}_{\rm sa}-\boldsymbol{J}_{\rm sp}-\boldsymbol{J}_{\rm sf},-\boldsymbol{M}_{\rm sf}\right)\right]^{\rm tan}\\&=0\qquad \text{on}\ S_{\rm p}\cup S_{\rm gf}\end{aligned} \tag{45}$$

$$\begin{aligned}&\left[\mathcal{E}_{\rm s}\left(\boldsymbol{J}_{\rm sa}+\boldsymbol{J}_{\rm sp}+\boldsymbol{J}_{\rm sg}+\boldsymbol{J}_{\rm sf},\boldsymbol{M}_{\rm sf}\right)\right]^{\rm tan}\\&=0\qquad \text{on}\ S_{\rm sa}\cup S_{\rm sp}\cup S_{\rm sg}\end{aligned} \tag{46}$$

$$\begin{aligned}&\left[\mathcal{E}_0\left(\boldsymbol{J}_{\rm a}+\boldsymbol{J}_{\rm p}+\boldsymbol{J}_{\rm gf}-\boldsymbol{J}_{\rm sa}-\boldsymbol{J}_{\rm sp}-\boldsymbol{J}_{\rm sf},-\boldsymbol{M}_{\rm sf}\right)\right]_{S_{\rm sf}^+}^{\rm tan}\\&=\left[\mathcal{E}_{\rm s}\left(\boldsymbol{J}_{\rm sa}+\boldsymbol{J}_{\rm sp}+\boldsymbol{J}_{\rm sg}+\boldsymbol{J}_{\rm sf},\boldsymbol{M}_{\rm sf}\right)\right]_{S_{\rm sf}^-}^{\rm tan}\ \text{on}\ S_{\rm sf}\end{aligned} \tag{47}$$

$$\begin{aligned}&\left[\mathcal{H}_0\left(\boldsymbol{J}_{\rm a}+\boldsymbol{J}_{\rm p}+\boldsymbol{J}_{\rm gf}-\boldsymbol{J}_{\rm sa}-\boldsymbol{J}_{\rm sp}-\boldsymbol{J}_{\rm sf},-\boldsymbol{M}_{\rm sf}\right)\right]_{S_{\rm sf}^+}^{\rm tan}\\&=\left[\mathcal{H}_{\rm s}\left(\boldsymbol{J}_{\rm sa}+\boldsymbol{J}_{\rm sp}+\boldsymbol{J}_{\rm sg}+\boldsymbol{J}_{\rm sf},\boldsymbol{M}_{\rm sf}\right)\right]_{S_{\rm sf}^-}^{\rm tan}\ \text{on}\ S_{\rm sf}\end{aligned} \tag{48}$$

where current $\boldsymbol{J}_{\rm sg}$ is the equivalent surface electric current on the interface between $V_{\rm s}$ and $V_{\rm g}$, operators $\mathcal{E}_{\rm s}$ and $\mathcal{H}_{\rm s}$ are with parameters $(\boldsymbol{\varepsilon}_{\rm s}^{\rm c},\mu_0)$, and the other detailed formulas related to Sec. IV-B are similar to the ones in Apps. B, D, and E.

## ACKNOWLEDGMENT

The authors would like to thank the reviewers and editors for their patient reviews, valuable comments, and selfless suggestions for improving this paper.